\documentstyle[epsfig, natbib209]{mn}

\newif\ifAMStwofonts

\ifoldfss
  \newcommand{\rmn}[1] {{\rm #1}}

  \ifCUPmtlplainloaded \else
    \NewTextAlphabet{textbfit} {cmbxti10} {}
    \NewTextAlphabet{textbfss} {cmssbx10} {}
    \NewMathAlphabet{mathbfit} {cmbxti10} {} 
    \NewMathAlphabet{mathbfss} {cmssbx10} {} 
  \fi
  \ifAMStwofonts
    \ifCUPmtlplainloaded \else
      \NewSymbolFont{upmath} {eurm10}
      \NewSymbolFont{AMSa} {msam10}
      \NewMathSymbol{\upi}     {0}{upmath}{19}
      \NewMathSymbol{\umu}     {0}{upmath}{16}
      \NewMathSymbol{\upartial}{0}{upmath}{40}
      \NewMathSymbol{\leqslant}{3}{AMSa}{36}
      \NewMathSymbol{\geqslant}{3}{AMSa}{3E}

       \let\le=\leqslant
       
    \fi
  \fi
\fi 

\ifnfssone
  \newmathalphabet{\mathit}
  \addtoversion{normal}{\mathit}{cmr}{m}{it}
  \addtoversion{bold}{\mathit}{cmr}{bx}{it}
  \newcommand{\rmn}[1] {\mathrm{#1}}

  \newmathalphabet{\mathbfit} 
  \addtoversion{normal}{\mathbfit}{cmr}{bx}{it}
  \addtoversion{bold}{\mathbfit}{cmr}{bx}{it}
  \newmathalphabet{\mathbfss} 
  \addtoversion{normal}{\mathbfss}{cmss}{bx}{n}
  \addtoversion{bold}{\mathbfss}{cmss}{bx}{n}
  \ifAMStwofonts
    \ifCUPmtlplainloaded \else
      \UseAMStwoboldmath
      \makeatletter
      \new@mathgroup\upmath@group
      \define@mathgroup\mv@normal\upmath@group{eur}{m}{n}
      \define@mathgroup\mv@bold\upmath@group{eur}{b}{n}
      \edef\UPM{\hexnumber\upmath@group}
      \new@mathgroup\amsa@group
      \define@mathgroup\mv@normal\amsa@group{msa}{m}{n}
      \define@mathgroup\mv@bold\amsa@group{msa}{m}{n}
      \edef\AMSa{\hexnumber\amsa@group}
      \makeatother
      \mathchardef\upi="0\UPM19
      \mathchardef\umu="0\UPM16
      \mathchardef\upartial="0\UPM40
      \mathchardef\leqslant="3\AMSa36
      \mathchardef\geqslant="3\AMSa3E

       \let\le=\leqslant

    \fi
  \fi
\fi 

\ifnfsstwo
  \newcommand{\rmn}[1] {\mathrm{#1}}

  \DeclareMathAlphabet{\mathbfit}{OT1}{cmr}{bx}{it}
  \SetMathAlphabet\mathbfit{bold}{OT1}{cmr}{bx}{it}
  \DeclareMathAlphabet{\mathbfss}{OT1}{cmss}{bx}{n}
  \SetMathAlphabet\mathbfss{bold}{OT1}{cmss}{bx}{n}
  \ifAMStwofonts
    \ifCUPmtlplainloaded \else
      \DeclareSymbolFont{UPM}{U}{eur}{m}{n}
      \SetSymbolFont{UPM}{bold}{U}{eur}{b}{n}
      \DeclareSymbolFont{AMSa}{U}{msa}{m}{n}
      \DeclareMathSymbol{\upi}{0}{UPM}{"19}
      \DeclareMathSymbol{\umu}{0}{UPM}{"16}
      \DeclareMathSymbol{\upartial}{0}{UPM}{"40}
      \DeclareMathSymbol{\leqslant}{3}{AMSa}{"36}
      \DeclareMathSymbol{\geqslant}{3}{AMSa}{"3E}

       \let\le=\leqslant

    \fi
  \fi
\fi 

\ifCUPmtlplainloaded \else
  \ifAMStwofonts \else 
    \def\upi{\pi}
    \def\umu{\mu}
    \def\upartial{\partial}
  \fi
\fi

\title{Rotation speed of the first stars}
\author[A. Stacy, V. Bromm and A. Loeb]
       {Athena Stacy$^{1}$\thanks{E-mail: minerva@astro.as.utexas.edu}, Volker Bromm$^{1}$ and Abraham Loeb$^{2}$ \\
 $^{1}$Department of Astronomy and Texas Cosmology Center, University of Texas, Austin, TX 78712, USA \\
 $^{2}$Astronomy Department, Harvard University, 60 Garden Street, Cambridge, MA 02138, USA}

\pagerange{\pageref{firstpage}--\pageref{lastpage}}
\pubyear{2010}

\begin{document}

\maketitle
\topmargin-1cm

\label{firstpage}

\begin{abstract}
We estimate the rotation speed of Population III (Pop III) stars within a minihalo at $z \sim 20$ using a smoothed particle hydrodynamics (SPH) simulation, beginning from cosmological initial conditions. We follow the evolution of the primordial gas up to densities of 10$^{12}$ cm$^{-3}$. Representing the growing hydrostatic cores with accreting sink particles, we measure the velocities and angular momenta of all particles that fall onto these protostellar regions.  This allows us to record the angular momentum of the sinks and estimate the rotational velocity of the Pop~III stars expected to form within them. The rotation rate has important implications for the evolution of the star, the fate encountered at the end of its life, and the potential for triggering a gamma-ray burst (GRB).  We find that there is sufficient angular momentum to yield rapidly rotating stars ($\ga$ 1000 km s$^{-1}$, or near break-up speeds). This indicates that Pop~III stars likely experienced strong rotational mixing, impacting their structure and nucleosynthetic yields. A subset of them was also likely to result in hypernova explosions, and possibly GRBs.     
\end{abstract}

\begin{keywords}
cosmology: theory -- early Universe -- galaxies: formation -- stars: formation.
\end{keywords}

\section{Introduction}
The first stars, also known as Population III (Pop III) stars, are
believed to be early drivers of cosmic evolution
(e.g. \citealt{barkana&loeb2001,bromm&larson2004,ciardi&ferrara2005,glover2005,byhm2009,loeb2010}).
These stars are thought to have formed around $z\sim 20$ within
minihaloes of mass $M\sim 10^6$\,M$_{\odot}$
(e.g. \citealt{haimanetal1996,tegmarketal1997,yahs2003}).  Not only
did the radiation from the first stars likely start the process of
reionizing the intergalactic medium (IGM; e.g.
\citealt{kitayamaetal2004,syahs2004,whalenetal2004,alvarezetal2006,
johnsongreif&bromm2007}), but when some of these stars produced supernovae
(SNe) explosions, they released the first heavy elements into the IGM,
providing its initial metal enrichment
(e.g. \citealt{madauferrara&rees2001,moriferrara&madau2002,brommyoshida&hernquist2003,wada&venkatesan2003,normanetal2004,tfs2007,
greifetal2007,greifetal2010,wise&abel2008}).

The mass of the first stars is the main factor in determining their
cosmological impact.  Pop~III stars are generally believed to be very
massive ($\sim 100$ M$_{\odot}$;
e.g. \citealt{abeletal2002,brommetal2002}), though recent evidence for
fragmentation in primordial gas may imply that the typical Pop~III
mass was somewhat lower (\citealt{clarketal2008,
clarketal2010,turketal2009,stacyetal2010}). The stellar luminosity and
ionizing photon production primarily depend on mass, as does the end
state of the star. For instance, stars between 40 M$_{\odot}$ and 140
M$_{\odot}$ are expected to collapse directly into black holes, while
stars in the mass range of
140~M$_{\odot}$~$<$~$M_{*}$~$<$~260~M$_{\odot}$ will die as
pair-instability supernovae (PISNe; \citealt{heger&woosley2002}).
Below 40 M$_{\odot}$, stars are again expected to explode as
core-collapse SNe, leaving behind a neutron star or black hole.
\cite{nomotoetal2003}, however, find that the nature of the explosions
from this mass range may vary depending on the angular momentum of the
collapsing core.  Stars with little angular momentum will explode as
faint SNe, while stars of the same mass but higher angular momentum
will become extremely energetic hypernovae.

Pop III stars also have the potential to produce gamma-ray bursts
(GRBs), particularly given the connection between long-duration GRBs
and the deaths of massive stars (see
\citealt{woosley&bloom2006}). GRBs may provide one of the most
promising methods of directly probing the final stages of Pop~III
stars, provided they occurred with a high enough frequency
(e.g. \citealt{bromm&loeb2002,bromm&loeb2006,gouetal2004,belczynskietal2007}). 
 \cite{naoz&bromberg2007} used early {\it Swift} data and an idealized star formation 
rate model to estimate that Pop III stars may indeed produce GRBs 
at an efficiency of $\sim 10^{-4}$ GRBs per solar mass incorporated in primordial stars. 
For the collapsar model of GRB generation to operate, this will require sufficient
angular momentum in the Pop~III progenitor for an accretion torus to
form around the remnant black hole
(e.g. \citealt{woosley1993,lee&ramirez2006}).  The progenitor star
must also lose its hydrogen envelope to enable the relativistic jet to
penetrate through and exit the star (e.g. \citealt{zhangetal2004}).
Fulfilling both of these conditions can be difficult for a single-star
progenitor, however, because removing the extended hydrogen envelope
will also lead to removal of angular momentum in the core
(e.g. \citealt{spruit2002,hegeretal2005,petrovicetal2005}). These
conditions for a GRB may be more easily met, however, in a close
binary system that experiences Roche lobe overflow
(e.g. \citealt{leeetal2002,izzardetal2004}).  
 Let us also note an alternate scenario recently explored by
\cite{suwa&ioka2010}.  They analytically find that the jet can break out even from 
an extended hydrogen envelope of a Pop III star if the jet is powered by magnetic 
fields.  This interesting result warrants further numerical study.

Another possibility arises if a Pop III star has a large enough
spin. This can affect its nucleosynthesis and change the evolution off
the main sequence (MS), opening a new pathway for the formation of
single-star progenitor GRBs (e.g. \citealt{yoon&langer2005,
woosley&heger2006}; Ekstr{\"o}m et al. 2008a\nocite{ekstrometal2008a}).  
\cite{woosley&heger2006} find
through their stellar evolution models that very massive $\sim$ 20
M$_{\odot}$ stars rapidly rotating at $\simeq 400$ km s$^{-1}$
($\simeq$ 40\% of the break-up speed) can completely mix while on the
MS, bypassing the red giant phase and becoming a Wolf-Rayet (WR)
star. This evolutionary path may furthermore allow the star to retain
enough angular momentum to become a GRB, particularly if the star has
low-metallicity and thus experiences significantly reduced mass loss
compared to solar-metallicity WR stars. \cite{yoon&langer2005} agree,
using a different numerical methodology, that rotationally induced
mixing will allow a low-metallicity massive star to evolve into a
rapidly rotating WR star and potentially a GRB.  Finally,
Ekstr{\"o}m et al. (2008a) \nocite{ekstrometal2008a} studied the evolution of metal-free stars with
a range of masses (15-200 M$_{\odot}$) and a high rotation rate of 800
km s$^{-1}$, corresponding to a fraction of 40-70\% of their break-up
speed.  In contrast to the previous studies, in their models chemical
mixing was usually not sufficient for the red giant phase to be
avoided. In fact, they found that the rotating stars generally end
their lives at a cooler location of the Hertzsprung-Russel diagram
(HRD). In addition, rotating stars produced a higher amount of metals,
compared to their non-rotating counterparts. Ekstr{\"o}m et al. (2008a) \nocite{ekstrometal2008a}
attribute this difference to the fact that, unlike the earlier
studies, they did not include the magnetic dynamo mechanism of
\cite{spruit2002}. With or without this mechanism, however, all
studies conclude that stellar rotation altered the evolution and fate
of  low-metallicity and Pop III stars. 
We also point out that, though we sometimes refer to low-metallicity studies,  
Pop III evolution is distinct from that of low-metallicity, and results for one 
do not simply extrapolate to the other (e.g. Ekstr{\"o}m et al. 2008b\nocite{ekstrometal2008b}). This 
highlights the need for continued investigation of rotating metal-free stars.

It is apparent that the angular momentum of Pop~III stars plays a key role in their evolution and death, as well as
their subsequent impact on the IGM. Whereas the mass scale of the
first stars has been investigated in numerous studies, their spin
remains poorly understood. It is an open question whether Pop~III
stars can realistically attain the high spin needed for the
above-mentioned processes to occur. Current observations of massive O
and B-type stars in our Galaxy and the Large Magellanic Cloud show
that they can indeed be rapid rotators, spinning at a significant
percentage of break-up speed (a few tens of percent).  They have a
large range of spin, from several tens of km s$^{-1}$ to well over 300
km s$^{-1}$, with an average of about $100$--$200$\,km s$^{-1}$
(e.g. \citealt{huang&gies2008,wolffetal2008}).  This does not
necessarily apply to Pop III stars, however, which formed in a
different environment.  While Pop~III stars formed in minihaloes whose
gravitational potential wells were dominated by dark matter (DM),
massive stars today form within molecular clouds that are not DM
dominated, and are embedded in much larger galaxies. The angular
momentum in the latter case ultimately derives from galactic
differential rotation on the largest scales, and turbulence in the
interstellar medium (ISM) on smaller scales (see,
e.g. \citealt{bodenheimer1995}).  \cite{wolffetal2008} argue that
their measured stellar rotation rates reflect the initial conditions
of the cores in which the stars formed, particularly the core
turbulent speeds and resulting infall rates.  To better determine the
possible rotation rates of Pop III stars, it is thus necessary to
study the environment specific to where they formed.

To better understand the potential for the various spin-dependent
evolutionary pathways of Pop III stars, we perform a three-dimensional
cosmological simulation that follows the evolution of primordial gas
in a minihalo to densities of $n>10^{12}$ cm$^{-3}$.  Similar to
\cite{bromm&loeb2004} and \cite{stacyetal2010}, we represent
gravitationally collapsing high-density peaks using the sink particle
method first introduced by \cite{bateetal1995}.  This allows us to
follow the mass flow onto the sinks for many ($\sim$ 100) dynamical
times.  As a sink particle grows in mass, the angular momentum of the
accreted mass is recorded, allowing us to measure the total angular
momentum of the sink and estimate the spin of the Pop III star
represented by the sink.  This is very similar to the method used in
\cite{jappsen&klessen2004} when they studied the angular momentum
evolution of protostellar cores, though their calculation had lower
resolution and was designed to study modern-day star formation as seen
in the ISM.  We give further details concerning our numerical
methodology in \S 2, while in \S 3 we present our results,
including estimates of the stellar rotation rate.  In \S 4 we
discuss the implications for the evolution and death of Pop III stars,
and in \S 5 we address the possibility of sub-sink fragmentation.
We summarize our main conclusions in \S 6.
  
\begin{figure*}
\includegraphics[width=.45\textwidth]{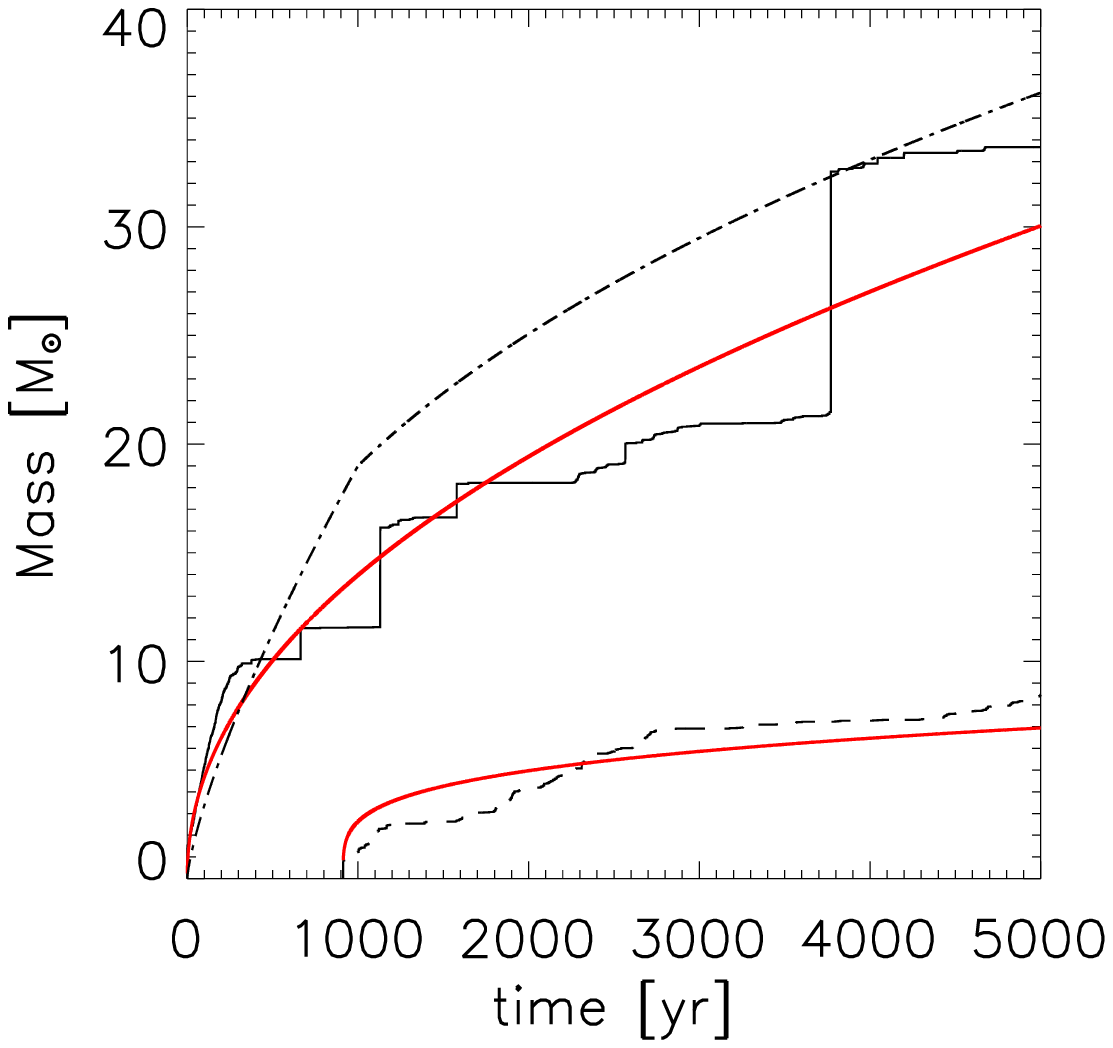}
\includegraphics[width=.45\textwidth]{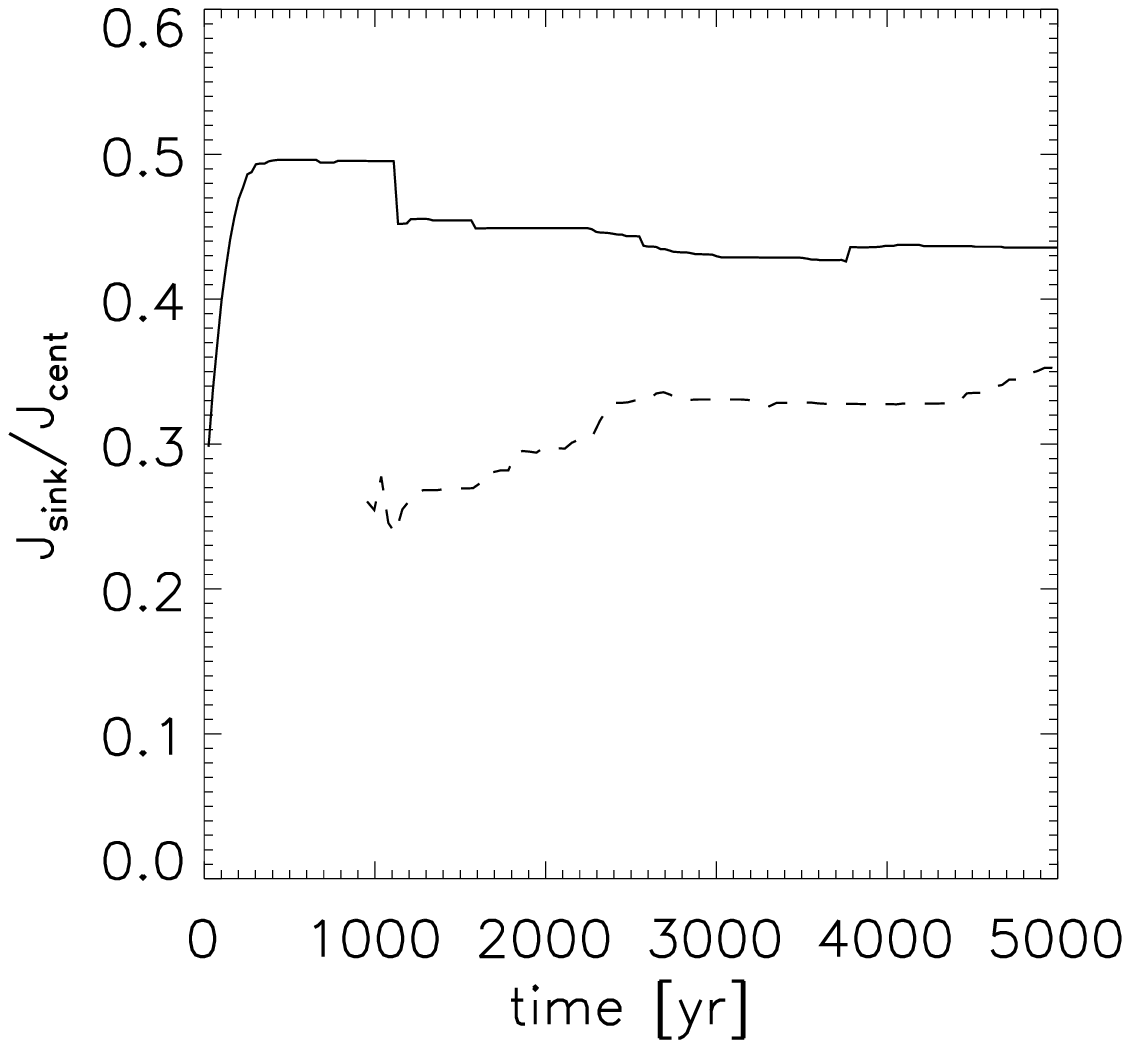}
\caption{{\it Left}: Growth of sink mass over time. Solid line is the growth of sink A, and dashed line is the growth of sink B.  Dash-dot line is the result from Bromm \& Loeb (2004). Red lines are power-law fits to the mass curves.  Though sink A grows rapidly for the first few hundred years, adding the accretion criterion of non-rotational support later causes its growth rate to be somewhat lower than that found in Bromm \& Loeb (2004) until a large merger event at 3800 years.
{\it Right}: Ratio $\epsilon = J_{\rmn sink}/J_{\rmn cent}$ as sinks grow over time.  Representation of different sinks is the same as in the left panel.  For sink B's accretion and the first $\sim$ 1000 years of sink A's accretion, note the similarity in how both mass and  $\epsilon$ increase over time.  This is due to the steadily growing rotational support of the mass that flows onto the sinks.}
\label{sink_mass}
\end{figure*}

\section{Numerical Methodology}

\subsection{Initial Setup}
Similar to the method used in \cite{stacyetal2010}, we carry out our
study using GADGET, a three-dimensional smoothed particle
hydrodynamics (SPH) code (\citealt{springeletal2001,
springel&hernquist2002}).  We perform the final part of the simulation
described in \cite{stacyetal2010} again, starting from approximately
4000 years ($\sim$ 100 free-fall times) before the first sink particle
forms.  This simulation was originally initialized at $z=99$ in a
periodic box of length 100 $h^{-1}$ kpc using both SPH and DM
particles.  This was done in accordance with a $\Lambda$CDM cosmology
with $\Omega_{\Lambda}=0.7$, $\Omega_{\rmn M}=0.3$, $\Omega_{\rmn
B}=0.04$, and $h=0.7$. To accelerate structure formation, we chose an
artificially enhanced normalization of the power spectrum of
$\sigma_8=1.4$. We have verified that the density and velocity fields
in the center of the minihalo are very similar to previous
simulations.  Even though we used an artificially high value of
$\sigma_8$, the angular momentum profile of our minihalo just before
sink formation was still very similar to that of other cosmological
simulations which used lower $\sigma_8$ values.  In particular, the
cosmological simulation of \cite{yoshidaetal2006}, which used
$\sigma_8 = 0.9$, and that of \cite{abeletal2002}, which used
$\sigma_8 = 0.7$, resulted in minihalo profiles which were especially
similar to ours on the smaller scales from which the mass of the sinks
is accreted. This demonstrates that our realization leads to
conditions that are typical for primordial star formation (see the
discussion in \citealt{stacyetal2010}).

To achieve high resolution we employed a standard hierarchical zoom-in procedure (see \citealt{stacyetal2010} for further details).  This involved adding three additional nested refinement levels of length 40, 30, and 20 kpc (comoving) centered on the site where the first minihalo will form.  Each level of higher refinement replaces particles from the lower level with eight child particles such that in the final simulation a parent particle is replaced by up to 512 child particles. The highest-resolution gas particles have a mass of $m_{\rmn SPH} =0.015$~M$_{\odot}$.  Therefore, the mass resolution of the refined simulation is: $M_{\rmn res}\simeq 1.5 N_{\rmn neigh} m_{\rmn SPH} \la  1$ M$_{\odot}$, where $N_{\rmn neigh}\simeq 32$ is the typical number of particles in the SPH smoothing kernel (e.g. \citealt{bate&burkert1997}).  The main difference between the current simulation and that described in \cite{stacyetal2010} is that we now record the angular momenta and velocities of the sink-accreted particles before they become incorporated into the sinks.  This allows us to track the total spin of the sink particles as they grow in mass.  
  
\subsection{Chemistry, heating, and cooling}

The chemistry, heating, and cooling of the primordial gas is treated similarly to that in earlier studies such as \cite{bromm&loeb2004}, \cite{yoshidaetal2006}, and \cite{stacyetal2010}.  We track the abundance evolution of the following species: H, H$^+$, H$^-$, H$_2$, H$_{2}^{+}$, He, He$^+$, He$^{++}$, e$^-$, and the deuterium species D, D$^+$, D$^-$, HD, and HD$^+$.  In the high-density disk that forms within the minihalo, H$_2$ is the dominant cooling agent, and although deuterium is unimportant for the thermal and chemical evolution of the gas at the late stages of collapse and accretion studied here, we include it for completeness.  We use the same chemical network and the same cooling and heating terms as used in \cite{stacyetal2010}.  This included accounting for modified physics at densities greater than $\simeq 10^8$ cm$^{-3}$: three-body processes which accelerate the formation of H$_2$ until the gas becomes fully molecular around $\simeq 10^{10}$ cm$^{-3}$, enhanced cooling due to collisions between H$_2$ molecules, H$_2$ formation heating, and modified values for the adiabatic exponent $\gamma_{\rmn ad}$ and the mean molecular weight $\mu$.  As described in \cite{stacyetal2010}, the evolution of the primordial gas up to the formation of the first sink particle was consistent with that of previous studies.    
 
\begin{figure*}
\includegraphics[width=.45\textwidth]{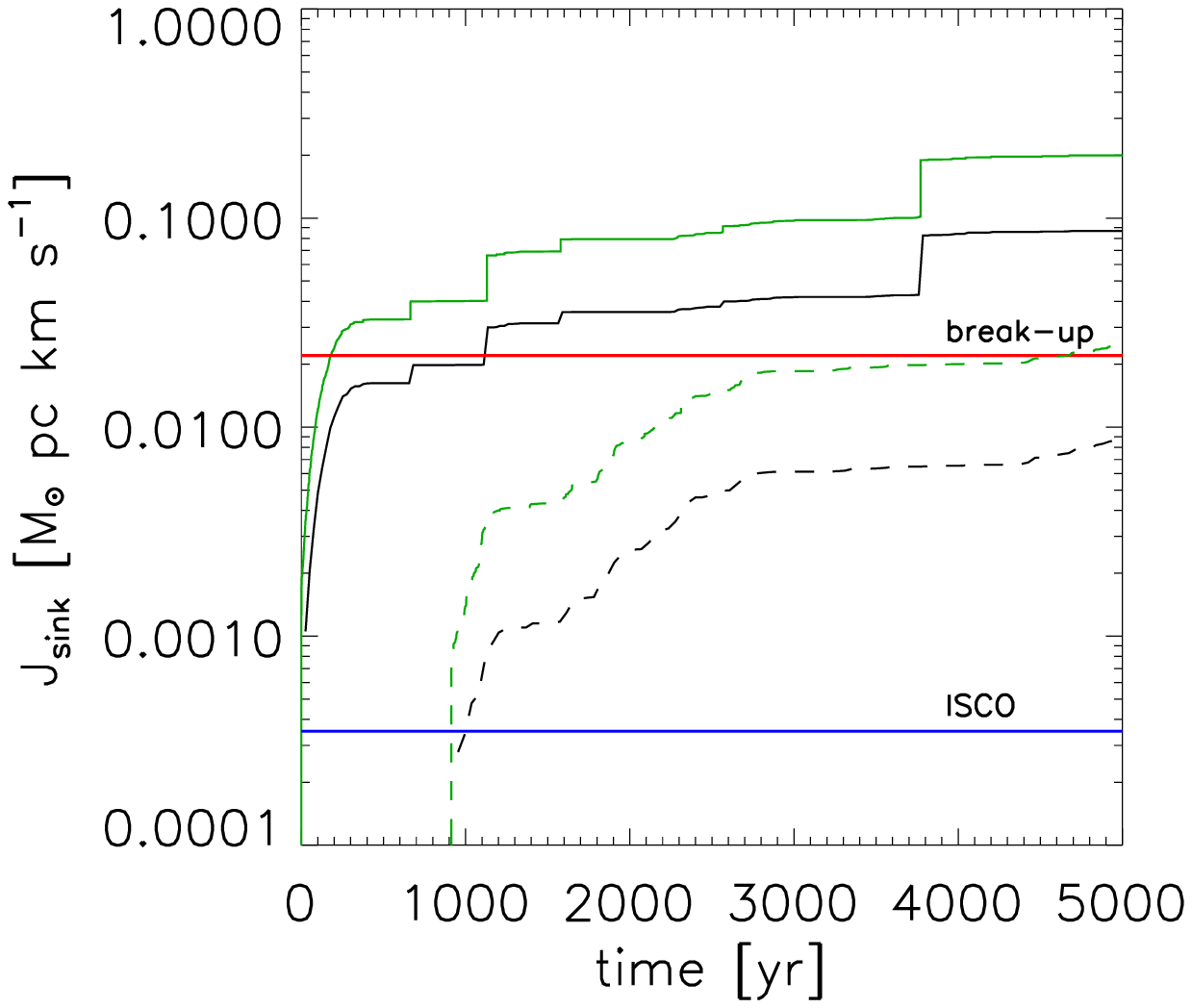}
\includegraphics[width=.45\textwidth]{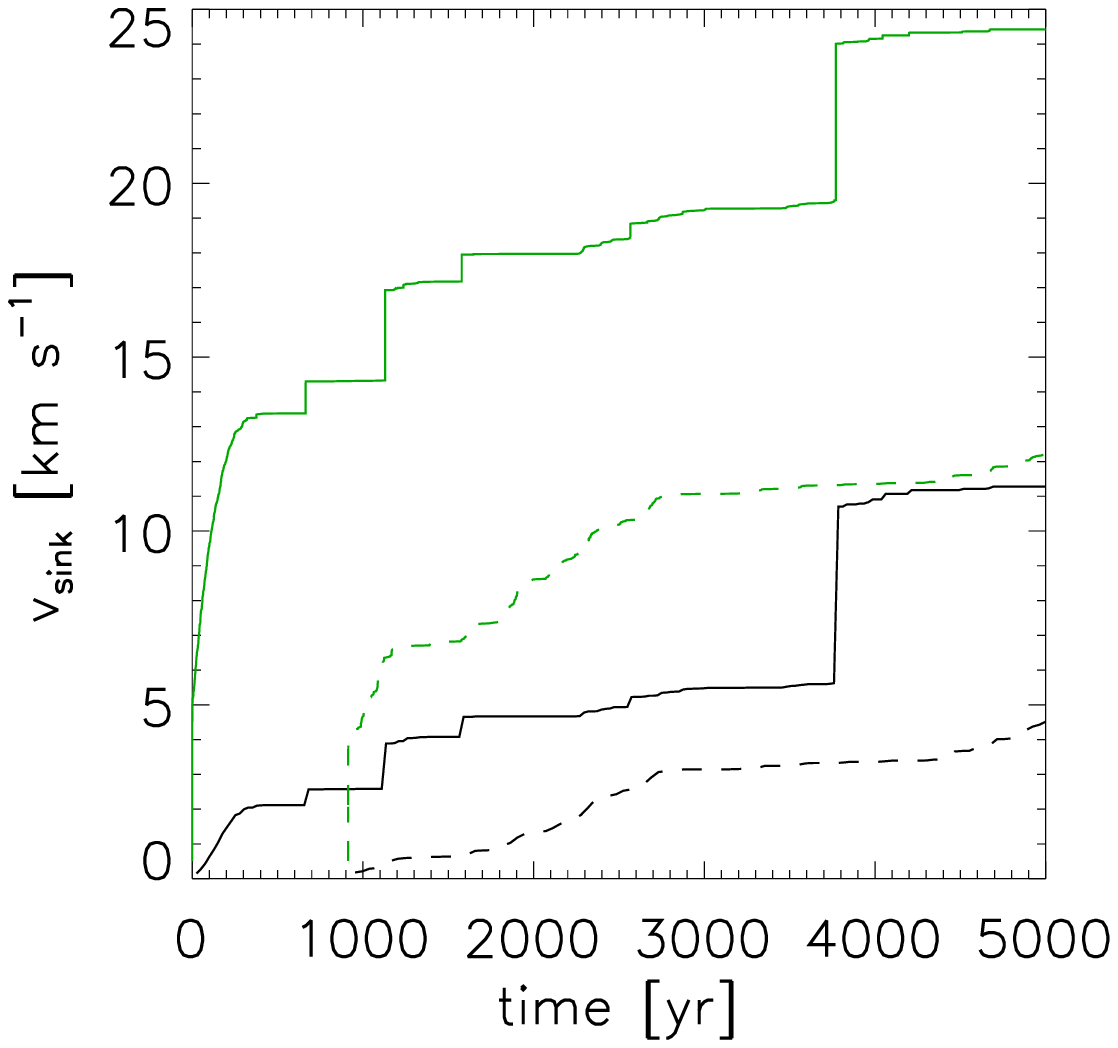}
\caption{{\it Left}: Angular momentum $J_{\rmn sink}$ of the sinks
over time, taken by summing the angular momentum $J_{\rmn SPH} =
m_{\rmn SPH}v_{\rmn rot}d$ of each accreted particle just before it is
added onto the sink.  Solid black line represents sink A, dashed black
line represents sink B.  The angular momentum of each sink grows as it
accretes more mass, always staying at a fraction of $J_{\rmn cent}$
(solid green line for sink A, dashed green line for sink B). Note that
for sink A, $J_{\rmn sink}$ exceeds the stellar $J_{\rmn break-up}$
(red line) by the end of the simulation.  The horizontal blue line
shows the orbital angular momentum at the {\it innermost stable
circular orbit (ISCO)}, $J_{\rmn ISCO}$, for a 100 M$_{\odot}$ black
hole. This value is well exceeded by $J_{\rmn sink}$ of both sinks.
{\it Right}: Sink rotational velocity, $v_{\rmn sink}$, measured at
$r=r_{\rmn acc}$ (with sink A denoted by a solid black line, and sink
B by a dashed black line). Values for $v_{\rmn sink}$ are recorded as
the mass-weighted average of $v_{\rmn rot}$ for all accreted
particles.  The upper (green) lines for each line type represent the
velocity needed for centrifugal support, $v_{\rmn cent}$.  Note that
$v_{\rmn sink}$ stays at a nearly constant fraction of $v_{\rmn
cent}$.  }
\label{angmom}
\end{figure*}

\subsection{Sink Particle Method}

We convert an SPH particle into a sink particle if it reaches a number
density of $n_{\rmn max} = 10^{12}$ cm$^{-3}$.  SPH particles that are
within a distance $r_{\rmn acc}$ of the sink are removed from the
simulation and their mass is also added to that of the sink, provided
that they are not rotationally supported against infall towards the
sink.  We set $r_{\rmn acc}$ equal to the resolution length of the
simulation, $r_{\rmn acc} = L_{\rmn res} \simeq 50$\,AU, where:
\begin{displaymath}
L_{\rmn res}\simeq 0.5 \left(\frac{M_{\rmn res}}{\rho_{\rmn max}}\right)^{1/3} \mbox{\ ,}
\end{displaymath}
with $\rho_{\rmn max}\simeq n_{\rmn max}m_{\rmn H}$ and $m_{\rmn H}$
being the proton mass.  The sink particle's mass, M$_{\rmn sink}$, is
initially close to the resolution mass of the simulation, $M_{\rmn
res} \simeq 0.7$ M$_{\odot}$.

We check for rotational support by comparing the specific angular
momentum of the SPH particle, $j_{\rmn SPH} = v_{\rmn rot} d$, with
the requirement for centrifugal support, $j_{\rmn cent} = \sqrt{G
M_{\rmn sink} r_{\rmn acc}}$, where $v_{\rmn rot}$ and $d$ are the
rotational velocity and distance of the particle relative to the
sink.  Once the sink is formed, any SPH particle that satisfies $d <
r_{\rmn acc}$ and $j_{\rmn SPH} < j_{\rmn cent}$ is accreted onto the
sink.  A sink particle can also be merged with another sink particle
if these same criteria are met.  When the sink is first formed, and
after each subsequent accretion event, its position and velocity are
set to the mass-weighted average of the particles it has accreted.  In
this way sink particles can grow and accrete mass over time.

As discussed in \cite{brommetal2002} and \cite{stacyetal2010}, our criteria for sink formation should be robust.  A gas particle must collapse two orders of magnitude above the average density of the surrounding disk, $\simeq 10^{10}$ cm$^{-3}$, before it is above the density threshold for sink formation.  This along with the small value for $r_{\rmn acc}$ and the further accretion criterion of non-rotational support ensures that sinks are indeed formed from gravitationally collapsing gas.  

Sink particles are held at a constant density of $n_{\rmn max}$ =
10$^{12}$ cm$^{-3}$, a constant temperature of 650 K, and a constant
pressure corresponding to its temperature and density.  Giving the
sink a temperature and pressure prevents the existence of a pressure
deficit around the sink that otherwise would yield an artificially high
accretion rate (see \citealt{brommetal2002,marteletal2006}). However,
the sink can still evolve in position and velocity due to
gravitational and hydrodynamical interactions.

The sink particle method is very useful for various reasons.  It
eliminates the need to incorporate chemistry, hydrodynamics and
radiative transfer at extremely high densities ($n>10^{12}$
cm$^{-3}$).  More importantly, by stopping the density growth at
$n>10^{12}$ cm$^{-3}$, the method allows the evolution of the region
of interest to be followed for many dynamical times. Without the sink
particle method, this would be computationally challenging because the
increasing density would lead to prohibitively small numerical
timesteps, a problem sometimes called `Courant myopia'.  Finally, the
sink particle method allows for a direct measurement of the angular
momentum growth and the accretion rate onto the high-density region
instead of having to indirectly extrapolate this from the
instantaneous density and velocity profiles at the end of the
simulation (e.g. \citealt{abeletal2002,yoshidaetal2006}).

\begin{figure*}
\includegraphics[width=.45\textwidth]{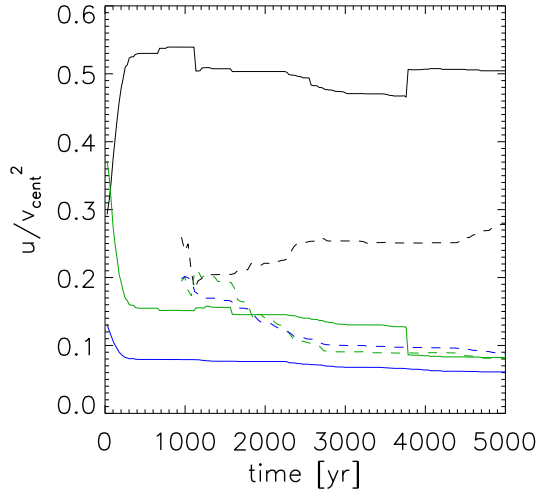}
\caption{Ratio of the energy of sink-accreted particles to the
specific gravitational energy of the sinks ($v^2_{\rmn cent}=GM_{\rmn
sink}/r_{\rmn acc}$).  Solid lines are for sink A, and dashed lines
are for sink B. Upper (black) lines of each line type represent the
energy of rotational motion, middle (green) lines represent the energy
of radial motion, and lower (blue) lines represent thermal energy.
For both sinks, the rotational energy component dominates for the
majority of the accretion time, and a thin Keplerian disk is likely to
develop on sub-sink scales. }
\label{egy_ratio}
\end{figure*}

\section{Results}

\subsection{Sink Growth and Angular Momentum}

\subsubsection{Accretion Rate}

The first minihalo within the cosmological box forms at $z \simeq 20$.
The subsequent evolution of the central region of the minihalo to
densities of $n = 10^{12}$ cm$^{-3}$ is described in
\cite{stacyetal2010}.  The growth of the first sink that forms, which
will be referred to as sink A, is similar to that found in
\cite{bromm&loeb2004} as well as \cite{stacyetal2010}.  The mass
growth is especially similar to that in \cite{stacyetal2010} for the
first few hundred years, up to several dynamical times after the sink
initially forms.  Though the sink accretion criteria have some small
differences in each of these three studies, this similarity in initial
sink growth points especially to the robustness of the density
threshold criterion for initial sink formation.  After 5000 years of
accretion, sink A grows to a mass of 34 M$_{\odot}$, similar to that
found in \cite{bromm&loeb2004}.  However, this is largely due to a
significant merger event at around 3800 years, and before this the
mass of sink A is around $1/3$ below that found in
\cite{bromm&loeb2004}.  Furthermore, the final sink mass is slightly
less ($\sim 20 \%$) than the final mass found in \cite{stacyetal2010}.
The reduced accretion rate found in this current calculation likely
arises because a sink is not allowed to accrete a particle if that
particle is rotationally supported against infall onto the sink, which
is an additional condition that was not included by
\cite{stacyetal2010}.  This condition seems to slightly decrease
the number of sink merger events, though the growth rate between
merger events is also somewhat reduced.

Around 300 years after the formation of sink A, a second sink forms.  Meanwhile, as sink A grows, a disk with radius of about 1000 AU develops around the sink, and disk fragmentation allows further sinks to form.  By the end of the simulation, 5000 years after sink A first forms, there is a total of four sinks.  The sink that is second-most massive, which we will label sink B, has grown almost to 9 M$_{\odot}$ (Fig. \ref{sink_mass}), while the remaining two sinks are $\sim$ 1 and 7 M$_{\odot}$.  
The overall accretion rate of sink B, $\simeq 2 \times 10^{-3}$ M$_{\odot}$ yr$^{-1}$, is around $30 \%$ that of sink A, $\simeq 7 \times 10^{-3}$ M$_{\odot}$ yr$^{-1}$.  The accretion rate for both sinks does not stay at a steady value, however, and actually declines as the sinks grow.  To show this we also provide power-law fits to the sink growth (red lines in Fig \ref{sink_mass}).  For sink A, $M_{\rmn sink} \propto t^{0.48}$, and $\dot{M} \propto t^{-0.52}$.  For sink B, $M_{\rmn sink} \propto t^{0.25}$, and $\dot{M} \propto t^{-0.75}$.

\subsubsection{Angular Momentum}

\begin{figure*}
\includegraphics[width=.45\textwidth]{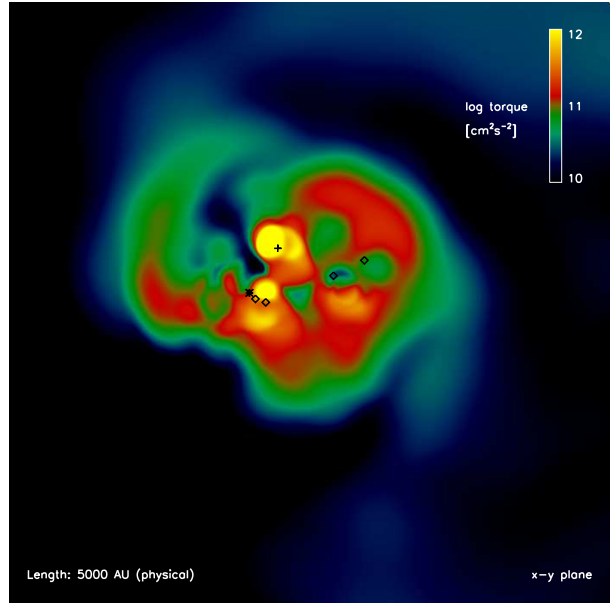}
\caption{Specific torque acting on the gas within the central 5000 AU of the simulation box, the region enclosing the large-scale star-forming disk.  Torques are calculated at a representative time of 2500 years after the first sink forms.  Shown is the z-component, perpendicular to disk plane, of all contributions to specific torque as measured from sink A.  The asterisk denotes the location of sink A, the cross denotes the location of sink B, and the diamonds are the locations of the remaining lower-mass sinks.  There is a total of six sinks, but this number will later be reduced through sink mergers. Note the spiral structure, where gravitational torques will remove angular momentum from the disk center on a timescale of approximately 100-1000 years.}
\label{torque}
\end{figure*}

\begin{figure*}
\includegraphics[width=.45\textwidth]{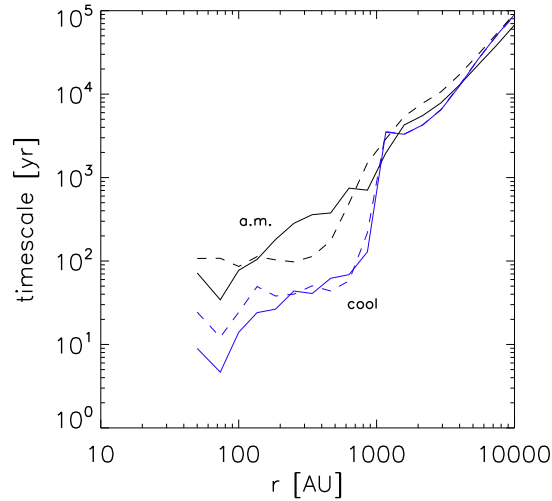}
\caption{Radially averaged timescales versus distance from the sink at
a typical accretion time of 2500 years.  The cooling timescale,
$t_{\rmn cool}$, is shown in blue as the lower set of lines, and the
angular momentum loss timescale, $t_{\rmn am}$, is shown in black as
the upper set of lines.  Solid lines are for sink A, and dashed lines
are for sink B.  For both sinks, $t_{\rmn cool}$ is an order of
magnitude shorter than $t_{\rmn am}$ from the sink edge out to $\sim$
1000 AU, the edge of the large-scale disk.}
\label{t_scales}
\end{figure*}

As the sinks grow in mass, the angular momentum of each particle accreted, $J_{\rmn SPH} = m_{\rmn SPH}v_{\rmn rot}d$, is added to the total angular momentum of the sink, $J_{\rmn sink}$.  Note that the scale of the sink, r$_{\rmn acc} = 50$ AU, is smaller than the radius corresponding to the sonic point:

\begin{equation}
r_{\rm sp} = \frac{G M_*}{c_s^2} \mbox{,}
\end{equation}

\noindent where $c_{\rmn s}$ is the sound speed. A typical sink mass
is 10 M$_{\odot}$, and the highest-temperature, high-density ($n >
10^8 \rmn cm^{-3}$) gas is approximately at 7000~K, or $c_{\rmn s}
\simeq 7$ km s$^{-1}$. This yields $r_{\rm sp} \simeq$ 180 AU.  For
the cooler disk gas, $T \sim 500$ K and $c_{\rmn s} \simeq 2$ km
s$^{-1}$, so that $r_{\rm sp}$ is larger, $\sim$ 2000 AU, and
similar to the size of the large-scale disk.  The sinks therefore
easily resolve the scale of the sonic point.  Angular momentum
transport from inside $r_{\rm sp}$ to outside $r_{\rm sp}$ will be
difficult once the inflow becomes supersonic
(\citealt{ulrich1976,tan&mckee2004}). In Section 3.2, we will show
that the inflow onto the sinks does indeed quickly become supersonic,
and in fact the total angular momentum within the sonic point steadily
grows as more mass continues to fall in. However, the angular momentum
inside $r_{\rm sp}$ can still be redistributed through torques within
the large-scale disk.
 
$J_{\rmn sink}$ stays at a fairly steady fraction, $\epsilon = J_{\rmn sink}/J_{\rmn cent}$, of the angular momentum required for full centrifugal support at the accretion radius,  $J_{\rmn cent} = M_{\rmn sink} j_{\rmn cent}$.  Figures \ref{sink_mass} and \ref{angmom} show this fraction to range from $\epsilon \simeq 0.45 - 0.5$ for sink A and $\epsilon \simeq 0.25 - 0.35$ for sink B.  Thus, on the scale of the accretion radius the sinks never become fully rotationally supported. Rotational velocities of the accreted particles were also recorded, and the total rotational velocity of each sink, $v_{\rmn sink}$, can be determined through a mass-weighted average of each accreted particle's $v_{\rmn rot}$ (right panel of Fig. \ref{angmom}).  Similar to the behavior of $J_{\rmn sink}$,  $v_{\rmn sink}$ stays at the same fairly constant fraction of  $v_{\rmn cent} =  \sqrt{G M_{\rmn sink}/r_{\rmn acc}}$.   

The specific angular momentum of the sinks does not stay perfectly
constant, however.  Comparing the mild evolution of $\epsilon$ with
the sinks' mass growth (Fig. \ref{sink_mass}) shows a general
correspondence between them, similar to that found in the simulations
of \cite{jappsen&klessen2004}.  For sink B in particular, the shape of
the mass versus time curve is very similar to that of the $\epsilon$
versus time curve, and the same applies for the first $\sim$ 1000
years of sink A's accretion.  These are periods when the mass that
flows onto the sinks is gradually increasing in rotational support as
the large-scale disk spins up.  The gas close to sink A is the first
to spin up, and so sink A reaches its maximum $\epsilon$ earlier on in the
simulation.

It is interesting to compare $J_{\rmn sink}$ with the minimum angular
momentum required for centrifugal support against infall onto a black
hole, as this is one of the minimum requirements for a successful
collapsar engine to power a GRB.  
For a  non-rotating black hole of mass $M_{\rmn
BH}$, the innermost stable circular orbit occurs at $r_{\rmn ISCO} =
6GM_{\rmn BH}/c^2$.  This corresponds to a minimum angular momentum of
$J_{\rmn ISCO} = \sqrt{6}GM_{\rmn BH}^2/c$,  which would be slightly smaller for rotating black holes.  
As can be seen in
Fig. \ref{angmom}, sink A and sink B both gather at least an order of
magnitude more angular momentum than necessary for a collapsar engine.
Whether this large sink-scale angular momentum continues down to
stellar scales is discussed below.

\subsection{Stellar Rotational Velocity}
\subsubsection{Thin Accretion Disk}

We now address how the measured sink spin can be extrapolated to the
scale of the final (MS) Pop~III star. To this end, let us compare
$J_{\rmn sink}$ to angular momentum values corresponding to rotational
break-up speeds on stellar scales.  A representative value for
$J_{\rmn break-up}$ can be found assuming a mass of 100 M$_{\odot}$
and a radius of 5 R$_{\odot}$, typical for massive Pop~III MS stars
(e.g. \citealt{brommetal2001}), 
 though this value may be somewhat larger for high rotation rates.  
In the latter half of the simulation,
$J_{\rmn sink}$ for sink B approaches $J_{\rmn break-up}$, while sink
A easily surpasses $J_{\rmn break-up}$ (red line in
Fig. \ref{angmom}). If all of sink A's angular momentum became
confined to smaller stellar scales this would thus be unphysical.
This becomes even more apparent after an analogous examination of sink
A's rotational velocity.  At the end of the simulation, sink A has a
rotational velocity of $v_{\rmn sink} = 11$ km s$^{-1}$.  We can
extrapolate $v_{\rmn sink}$ to the stellar scale of 5~R$_{\odot}$ by
assuming conservation of angular momentum to find $v_{*} = v_{\rmn
sink} r_{\rmn acc}/5 \rmn R_{\odot} $.  This turns out to be over
20,000~km~s$^{-1}$, significantly greater than our typical stellar
break-up velocity\footnote[2]{ The formally correct equation for
break-up velocity is $v_{\rmn break-up} = \sqrt{\frac{2}{3}G \, M/ \rmn R}$, where the factor of $\frac{2}{3}$ 
accounts for deformation due to rotation.  However, due to the approximate nature of our calculations, for 
simplicity we omit this factor of  $\frac{2}{3}$ from our calculations.}, $v_{\rmn break-up} \simeq \sqrt{G \, 100 \rmn M_{\odot}/
5 \rmn R_{\odot}} \simeq$ 2000 km s$^{-1}$ 
Again, this is unphysical, and serves as an example of the classic angular momentum
problem in the context of star formation (e.g. \citealt{spitzer1978,
bodenheimer1995}), now extended to the immediate protostellar
environment.

Further insight can be found by evaluating the centrifugal radius, 

\begin{equation}
r_{\rmn cent} = \frac{j_{\rmn sink}^2}{GM_{\rmn sink}} \mbox{,}
\end{equation}

\noindent where $j_{\rmn sink}=J_{\rmn sink}/M_{\rmn sink}$. For sink A, in the latter part of the simulation $j_{\rmn sink}$ is typically around $8 \times 10^{20}$~cm$^2$~s$^{-1}$, while sink~B has $j_{\rmn sink} \simeq 3 \times 10^{20}$~cm$^2$~s$^{-1}$.  At the end of the calculation, $r_{\rmn cent} \simeq 10 \, \rmn AU$ for sink A and $r_{\rmn cent} \simeq 6 \, \rmn AU$ for sink B, around two orders of magnitude larger than the stellar sizes cited above.  At a length sale of $r_{\rmn cent}$, contraction would be halted by the centrifugal barrier, and a disk would form.  Accretion onto the star would then continue through the disk, and the disk is expected to grow in size.  As mentioned above and described in \cite{stacyetal2010}, in the simulation a disk structure does indeed form and grow well beyond the sink radius.  Other very high-resolution simulations also find that primordial gas develops into disks on small scales, less then 50 AU from the star (\citealt{clarketal2008,clarketal2010}).  

We therefore infer that much of the angular momentum of the sinks will
be distributed in a disk, while most of the sink mass lies within the
small $\simeq 5 \, \rmn R_{\odot}$ star.  There is evidence for the
existence of similar disk structure around massive stars in the Galaxy
(see, e.g. \citealt{cesaronietal2006, krausetal2010}).  The nature of
the disk can be estimated through a comparison of the thermal energy
with the kinetic energy of rotational and radial motion at the sink
accretion radius.  For gas flow onto a gravitationally dominant
central mass, dimensionally the sum of these energies per unit mass,
should follow the approximate relation
(e.g. \citealt{narayan&yi1994}),
\begin{equation}
v_{\rmn rot}^2 + v_{\rmn rad}^2 + c_{\rmn s}^2 \sim \frac{G \,
M_{\rmn sink}}{r_{\rmn acc}} \equiv v_{\rmn cent}^2 \mbox{.}
\end{equation}
Since sink A is the dominant mass, the above relation will
more accurately apply to sink A than to sink B, but the energy
comparison remains useful for both.  Figure \ref{egy_ratio} shows these
energies for each sink relative to its specific gravity, or $v_{\rmn
cent}^2$, and how these ratios evolve over time.  Overall sink ratios
were calculated using a mass-weighted average over the individual
particles accreted.  For sink A, the rotational energy strongly
dominates after $\sim$ 300 years and stays dominant for the rest of
the simulation.  For sink B, the thermal energy and energy of radial
motion remain at similarly low values throughout most of the sink's
accretion.  Around 500 years after sink B forms, the rotational energy
becomes the largest contribution to the total sink energy, and this
dominance steadily grows for the rest of the calculation.  This
relatively large amount of rotational energy and low amount of thermal
and radial energy for both sinks implies that their sub-sink disks
will become thin and Keplerian.

A comparison of the cooling time, $t_{\rmn cool}$, with the timescale
for angular momentum loss, $t_{\rmn am}$, of the gas around the sinks
gives further supporting evidence for sub-sink, thin, Keplerian disks.
We calculate $t_{\rmn am}$ directly from the simulation by recording
the acceleration on each gas particle and determining the torque due
to numerical viscosity ($\vec\tau_{\rmn visc}$) as well as the torque
exerted by gravity and pressure ($\vec\tau_{\rmn grav}$ and
$\vec\tau_{\rmn pres}$).  The total torque on a given particle within
a gas cloud is given by
\begin{eqnarray}
\vec\tau_{\rmn  tot}&&=  \vec\tau_{\rmn grav} + \vec\tau_{\rmn pres}  + \vec\tau_{\rmn visc} \nonumber\\
&&= m_{\rmn SPH}\vec{d}\times(\vec{a}_{\rmn grav} + \vec{a}_{\rmn pres} + \vec{a}_{\rmn visc}) \mbox{\ ,}
\end{eqnarray}
where
\begin{eqnarray}
t_{\rmn am} \simeq J_{\rmn SPH}/|\vec\tau_{\rmn  tot}| \mbox{\ ,}
\end{eqnarray} 
and
\begin{eqnarray}
t_{\rmn cool} \simeq \frac{n k_{\rmn B} T}{\Lambda} \mbox{\ ,} 
\end{eqnarray}
with $k_{\rmn B}$ being the Boltzmann constant, $T$ the gas
temperature, and $\Lambda$ the cooling rate (in erg cm$^{-3}$
s$^{-1}$). We find that $\vec\tau_{\rmn grav}$ and $\vec\tau_{\rmn
pres}$ dominate, accounting for 80\% of the total.  Figure \ref{torque}
shows the $z$-component, perpendicular to the disk plane, of the
specific torque acting upon the gas within the large-scale disk, as
measured from the location of sink A.  The spiral structure indicates
the dominance of gravitational torques which remove angular momentum
from the center of the disk on timescales of $100-1000$ years,
enabling disk material to be accreted onto the sinks.

Fig. \ref{t_scales} shows these timescales for the gas particles in
radially averaged bins.  From this we can see that for the gas
surrounding each sink, $t_{\rmn cool}$ $\sim$ $t_{\rmn am}$ at
distances greater than 1000 AU.  However, at 1000 AU $t_{\rmn cool}$
falls below $t_{\rmn am}$.  This coincides well with the fact that
this is the radius of the large-scale disk which embeds the whole
stellar multiple system.  At the sink edges, $t_{\rmn cool}$ is nearly
an order of magnitude shorter than $t_{\rmn am}$ for both sinks.
Thermal energy of the gas is radiated away quickly enough that
rotational energy will likely remain dominant.  Though torques are
active, particularly gravitational ones from the spiral structure in
the disk, they are unlikely to remove angular momentum quickly enough
to prevent the formation of a sub-sink Keplerian disk once the central
stellar mass has grown substantially.

\subsubsection{Extrapolation to Stellar Surface}
If the entire extent of the sub-sink disks is indeed Keplerian and the
disk self-gravity is negligible, then gas within the sinks will rotate
at $v(r) \simeq v_{\rmn Kep}(r) \simeq \sqrt{G M_{\rmn *}/r}$, where
$r$ is the distance from the star, $M_* = f_* M_{\rmn sink}$ is the
mass of the star, and $f_*$ is the sink mass fraction that ends up in
the star while the remaining mass is stored in the disk. For $f_* \la
1$, we will have $v(r) \la \sqrt{G M_{\rmn sink}/r}$.  If the inner
edge of the disk extends all the way to the stellar surface, which is
expected if magnetic fields are not important (see \S 6), then the gas
acquired by the star from the accretion disk will be rotating at full
Keplerian velocity.


The location of the stellar surface varies as the star's radius evolves.  The total angular momentum acquired by the star will depend upon this evolution, and at any given time this total $J_*$ is given by  

\begin{eqnarray}
J_*(t) = \int_0^t \, j(R_*) \dot{M}\, dt = \int_0^t \, \sqrt{G M_{\rmn *}R_*} \dot{M}\, dt \mbox{\ ,}
\end{eqnarray}  

\noindent where $R_*$ is the radius of the star. To evaluate this expression, we use the same prescription for the protostellar radial evolution as described in \cite{stacyetal2010}, which in turn was based upon the earlier work of, e.g. \cite{stahler&palla1986} and \cite{omukai&palla2003}.  In this prescription, when the protostar first forms as a small hydrostatic core, it will initially undergo a phase of adiabatic accretion and gradual expansion.  During this time the protostellar radius will grow as

\begin{equation}
R_{*I} \simeq 50 {\rmn R_{\odot}} \left(\frac{M_*}{\rmn M_{\odot}}\right)^{1/3} \left(\frac{\dot{M}}{\dot{M}_{\rmn fid}}\right)^{1/3}   \mbox{\ ,}
\end{equation}

\noindent where $\dot{M}_{\rmn fid}\simeq 4.4\times 10^{-3} {\rmn M}_{\odot}$\,yr$^{-1}$ is a fiducial rate, typical for Pop~III accretion.
During the subsequent phase of Kelvin-Helmholtz (KH) contraction, the radius will shrink according to

\begin{equation}
R_{*II} \simeq 140 {\rmn R_{\odot}} \left(\frac{\dot{M}}{\dot{M}_{\rmn fid}}\right) \left(\frac{M_*}{10 \rmn M_{\odot}}\right)^{-2} \mbox{\ .}
\end{equation}

\noindent We estimate that the transition from adiabatic accretion to KH
contraction occurs when the value of $R_{*II}$ falls below that of
$R_{*I}$.  For $M_*$ and $\dot{M}$ we employ the power-law fits
discussed in \S 3.1, and we set $M_* \simeq
M_{\rmn sink}$ in the following analysis.  
In doing this we have made the simplifying assumption that nearly all of 
the gas accreted onto the sink quickly flows through the relatively low-mass disk onto the dominating massive star.
We also extend the fits to $10^5$ years, roughly the point
when KH contraction will cease and the star settles onto the MS, with
a final radius of 5\,R$_{\odot}$.  
Although the Pop~III radial evolution and MS size is based
on work that does not account for varying accretion rates and stellar rotation, which may
inflate the radius, this should still give a general picture of how the Pop~III rotational velocity will evolve.

\begin{figure*}
\includegraphics[width=.45\textwidth]{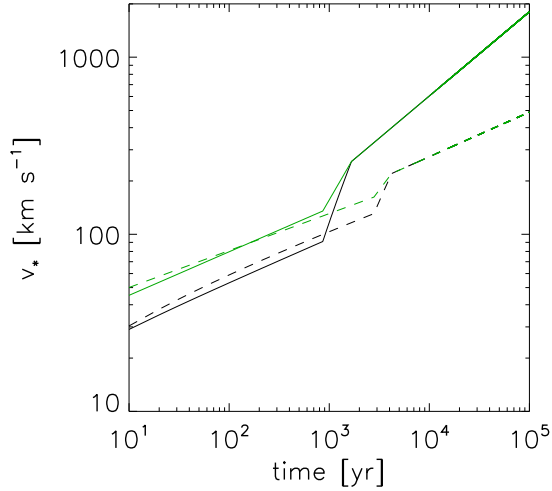}
\caption{Evolution of stellar rotation. Lower black lines show the
rotational velocities $v_{*}$ of the stars as they initially grow
slowly through adiabatic accretion and then undergo KH contraction
onto the MS.  Upper green lines are the break-up velocities of the
stars, $v_{\rmn max} \simeq \sqrt{G M_{\rmn sink}/R_*}$.  Solid
lines are for sink A, and dashed lines are for sink B.  Note that once
KH contraction begins at around 1000 years for sink A and 3000 years
for sink B, both stars quickly spin up to the full break-up
velocity. }

\label{vel}
\end{figure*}

The resulting evolution of
 $v_* = J_*/R_*$ for each sink is shown in Fig. \ref{vel}.  Note that during the
stars' initial slow expansion, the velocity is not quite at break-up
because the stars are gathering mass from gradually increasing radii.
Once the stars begin KH contraction, however, the total angular
momentum of the stars in fact exceeds break-up, but in this
case we assume that the angular momentum will slow the KH contraction
accordingly, and we adjust the stellar radius such that the star will
again rotate at break-up speed.  
By setting the right-hand side of Equ.~7 equal to $J_{\rmn Kep}$, we find that
the radius during this third slowed contraction phase will evolve according to 

\begin{equation}
\frac{d}{dt} {\rmn ln}\,R_{*III} = - \frac{d}{dt} {\rmn ln}\,M
\end{equation}

\noindent Once this phase begins, the star will
rotate at break-up speed, 
 $v_* = v_{\rmn max} \simeq \sqrt{G M_{\rmn
sink}/R_*}$.  

Given this model, at $10^5$ years the star within sink A
has mass of 125 M$_{\odot}$, a radius of 7 R$_{\odot}$, and a
rotational velocity of 1800 km s$^{-1}$.  The star within sink B has
mass of 15 M$_{\odot}$, a radius of 12 R$_{\odot}$, and a rotational
velocity of 500 km s$^{-1}$.  Though details of this model are
uncertain, particularly how the accretion rate will evolve during
later times beyond the end of our simulation, we can still expect
spin-up to occur during KH contraction, likely yielding rotational
velocities near break-up speed.

\begin{table*}
\begin{tabular}[width=.9\textwidth]{crrrrr}
\hline
sink & $M_*(5000 \, \rmn yr)$ [M$_{\odot}$]  & $M_*(10^5 \, \rmn yr)$ [M$_{\odot}$] & $j_{\rmn sink}$ [cm$^2$~s$^{-1}$]  & $v_*$ [km s$^{-1}$] & $v_{\rmn *,low}$ [km s$^{-1}$]\\
\hline
A  & 34  & 125 & $8 \times 10^{20}$  & 1800 & 800\\
B  & 9  &  15 & $3 \times 10^{20}$  & 500 & 300\\
\hline
\end{tabular}
\caption{Stellar masses at 5000 years, extrapolated mass at $10^5$ years, specific angular momenta of the sinks, final stellar rotational velocities $v_*$, and the more conservative estimate of stellar rotational velocity $v_{\rmn *,low}$. }
\label{tab1}
\end{table*}

We can also make a more conservative estimate to represent the
possible case that sub-sink torques do become strong enough to yield
sub-Keplerian rotation rates.  Since the overall sink angular momentum
stays at a fairly constant fraction of $J_{\rmn cent}$, we can apply
this to sub-sink scales as well.  Then we have $v(r) = \epsilon \,
v_{\rmn Kep}(r) = \epsilon \, \sqrt{G M_{\rmn sink}/r}$.  This is
similar to the situation described by \cite{narayan&yi1994} in which
the gas cannot cool efficiently, causing the accretion flow to stay at
approximately the virial temperature.  The resulting viscosity is high
relative to cold gas, allowing angular momentum to be transported
outwards and leading to rotational velocities that remain at a
constant fraction $\epsilon < 1$ of $v_{\rmn Kep}(r)$ for a large
range of radii.  At $10^5$ years, given $\epsilon = 0.45$ (see
Fig. \ref{sink_mass}), the star within sink A will be rotating at
 $v_{\rmn *,low} \simeq$ 800 km s$^{-1}$ (see Fig. \ref{vel}).  This is still a high
rotational velocity that is a substantial fraction of the break-up
speed.  The fastest-rotating stars considered by
\cite{woosley&heger2006} and \cite{yoon&langer2005}, for instance, had
$\epsilon$ values of 0.45~-~0.5.  The sink B star would not rotate
quite as rapidly due to its lower mass and its slightly lower values
of $\epsilon$, but for $\epsilon = 0.35$ the sink B star is still
estimated to reach a significant rotational velocity of  $v_{\rmn *,low} \simeq$ 300 km
s$^{-1}$.  Also note that, in the conservative case,  $v_{\rmn max}$  at 
$10^5$ years is slightly higher than that shown in Fig. \ref{vel} because KH contraction 
is no longer slowed by excess angular momentum, and the stars have already 
reached the MS radius by this time.

\begin{figure*}
\includegraphics[width=.45\textwidth]{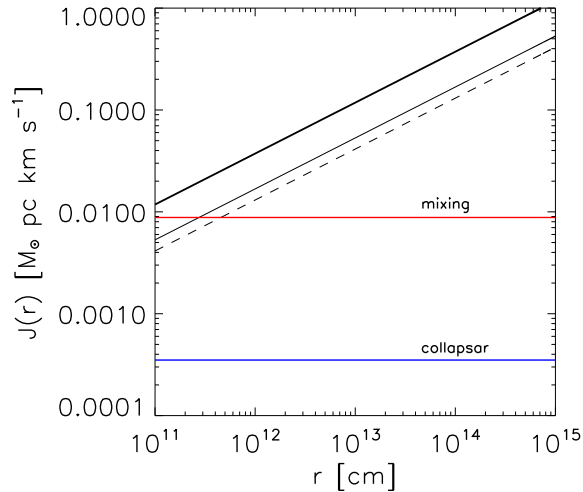}
\caption{Angular momentum relative to the center of the stars,
assuming they have grown to a mass of 100 M$_{\odot}$ and have an
approximate radius of 5 R$_{\odot}$. Thick black solid line represents
the situation of Keplerian rotation where $J(r) = J_{\rmn Kep}(r)$.
The thin diagonal lines represent the case of $J(r) = \epsilon \,
J_{\rmn Kep}(r)$.  For sink A (solid line), $\epsilon = 0.45$ was
used.  For sink B (dashed line), we used a smaller value of $\epsilon
= 0.35$. Blue line (labeled as ``collapsar'') shows $J_{\rmn ISCO}$.
Red line (labeled as ``mixing'') shows $0.4*J_{\rmn break-up}$, the
approximate minimum angular momentum necessary for a low-metallicity
star to undergo rotational mixing and chemically homogeneous
evolution, as determined by Yoon and Langer (2005) and Woosley and
Heger (2006).  The angular momentum requirement for the collapsar
engine, $J > J_{\rmn ISCO}$, is easily met on sub-sink
scales. Rotational mixing will readily occur as well as the stars
approach their break-up speed.}
\label{jvsr}
\end{figure*}

\section{Implications of Rapid Rotation}

\subsection{Rotational Mixing}

Numerous previous studies have found that high rotational velocities such as those predicted from our simulation will alter the stellar evolution (see, e.g. \citealt{maeder&meynet2000}).  Models of \cite{maeder1987}, for instance, find that above a critical velocity of 350 km s$^{-1}$ for 20 M$_{\odot}$ stars (or 30-40\% of break-up velocity), rotationally induced mixing will lead to a very different evolution.  Instead of the expected redward track off the MS in the HRD, mixing reduces the chemical gradient throughout the star enough that no extended hydrogen envelope forms. The star smoothly transitions from hydrogen to helium burning, and the stellar radius stays roughly constant while the temperature and luminosity both steadily increase.  The star then enters the WR stage, during which heavy mass loss leads to a decrease in the luminosity, though the temperature still remains high.  As will be further discussed in the next section, updated studies by \cite{yoon&langer2005} and \cite{woosley&heger2006} of low-metallicity stars, and also studies by \cite{heger&langer2000} of solar-metallicity stars, all similarly find that massive stars with high rotation rates ($\sim$ 40-50\% of break-up speed) can undergo chemically homogeneous evolution to become rapidly rotating WR stars.  

Rotational mixing is likely to occur according to the calculations we present here, particularly if the stars do indeed rotate at nearly full break-up speed (see Fig. \ref{jvsr}).  This will have important implications for Pop III feedback.  Effective temperatures of stars that undergo such rotational mixing can reach up to an order of magnitude higher than corresponding non-rotating stars, while luminosities may be two to three times as high (see, e.g. \citealt{yoon&langer2005}).  This will lead to an increased emission of ionizing radiation at harder wavelengths, so the HII regions will be larger than expected from non-rotating models of stars of the same mass (e.g. \citealt{greifetal2009}).     

It is important to note that these results do vary depending upon metallicity and details of the stellar model.  For instance, in contrast to earlier studies, Ekstr{\"o}m et al. (2008a) \nocite{ekstrometal2008a} find that even rotational speeds of up to 70\% of break-up speed will not be sufficient to drive chemically homogeneous evolution.  However, they do find rotation to increase the MS lifetime by 10-25\%, which again would increase the total amount of ionizing radiation from rotating massive Pop III stars.  A final distinction between rotating and non-rotating Pop III evolution is that rotating models generally yield higher total amounts of metals by the end of their nucleosynthesis (Ekstr{\"o}m et al. 2008a\nocite{ekstrometal2008a}).  Depending on how this metallicity gets spread to the star's surroundings through stellar winds and SN explosions, higher metallicity will enhance the later cooling and collapse of gas when subsequent generations of stars form, most likely lowering their average mass (e.g. \citealt{omukai2000,brommetal2001b,schneideretal2006,frebeletal2007,frebeletal2009,greifetal2010})   

\subsection{GRBs and Hypernovae}

The final fate of the stars in our simulation will depend on the mass
they reach through accretion, though computational limitations prevent
us from following the entire accretion history over the stellar
lifetime of $\sim$ 3 Myr.  However, extrapolating from the first 5000
years (see Fig. \ref{sink_mass}) implies that both stars we have
discussed are likely to grow significantly more massive.  They should
certainly be massive enough to avoid a white dwarf fate and make a
neutron star or black hole, assuming they do not die as PISNe and
leave behind no remnant at all.  If they in fact die as core-collapse
SNe, we can estimate the effect of rotation on the later SN explosion.
Though a black hole remnant is more likely, particularly for the more
massive star of sink A, we can derive a more conservative estimate by
considering a neutron star remnant.  As described in
\cite{woosley&heger2006}, the total rotational energy of a resulting
neutron star of radius 12 km and gravitational mass of 1.4 M$_{\odot}$
will be $E_{\rmn rot} \simeq 1.1 \times 10^{51} \, (5 \, {\rmn
ms}/P)^2$~erg, where $P$ is the rotation period of the neutron star.
They find that for $E_{\rmn rot}$ to be comparable to the energy of a
hypernova, $\sim 10^{52}$ erg, $P$ would need to be $\le$~2 ms.  In
their low-metallicity models that begin with stars rotating with
similar $\epsilon$ values to what we found for sink A ($\epsilon \sim
0.45)$, they infer resulting neutron star rotation rates that do meet
this criterion.  They find the same results even for lower $\epsilon$
like for those of sink B, though this is only if magnetic torques are
not included in the model.  Thus, even using the more conservative
estimate it is conceivable that the rotational energy reservoir found
in sink A and B of our work could be enough to power a hypernova.  If
the stars rotate more rapidly, at nearly their break-up velocity
as we predict, then hypernovae would be even more likely.

Will there still be enough angular momentum for the collapsar engine
to work on these stellar scales?  If we use the low estimate of $J(r)
= \epsilon \, J_{\rmn Kep}(r) = \epsilon \, \sqrt{G M_{\rmn sink}
\,r}$, then on these scales the angular momentum for both sinks still
easily meets the required $J > J_{\rmn ISCO}$, especially if their
high $\epsilon$ values continue as the stars grow to larger masses
(see Fig. \ref{jvsr}).  But will such stars retain this angular
momentum as they evolve?  Low metallicity stellar models in
\cite{yoon&langer2005} and \cite{woosley&heger2006} that were
initialized with $\epsilon$ values similar to that of sink A show that
such stars may indeed be able to retain sufficient amounts of angular
momentum in their cores throughout their evolution to the pre-SN
stage, depending upon the strength of magnetic fields and the mass
loss rate during the WR stage.  Though there is still some angular
momentum loss in their GRB-forming models, it is limited because
rotationally induced mixing allowed these stars to avoid a red giant
phase on their path to becoming WR stars.  However,
\cite{woosley&heger2006} generally find that when both magnetic fields
and strong mass loss are included in their models, the ability of even
their high-$\epsilon$ models to meet the GRB requirements becomes
borderline.

For the lower $\epsilon$ value of 0.35 as found for sink B, a GRB
becomes yet less likely, and the star cannot even go through a WR
phase unless the models exclude magnetic fields.  However, as
discussed in \S 3.2, the stars within both sink A and sink B are
expected to rotate at a much higher fraction of break-up speed - close
to 100\%.  In this case, the above path for becoming a GRB should
work yet more readily.  Unless there are mechanisms for significant
angular momentum transport away from the stars (see \S 6), the
angular momentum condition for the collapsar engine will be met.

\section{Sub-Sink Fragmentation}

Up to this point we have assumed that each sink will host a single
star-disk system, but it is possible that more than one star could
exist inside a sink.  For instance, a sink merger might lead to a
sub-sink binary instead of the presumed coalescence of two stars.  We
may also consider the possibility that the mass of the sink will
fragment and a sub-sink stellar multiple will form in this way.  As
described by \cite{jappsen&klessen2004}, this is more likely for
higher values of $\beta$, the ratio of rotational to gravitational
energy.  As in \cite{goodmanetal1993} and \cite{jappsen&klessen2004},
we can arrive at a simple estimate by assuming that the sinks are
undergoing solid-body rotation, have constant angular velocity
$\Omega$, and have uniform density.  In this case, sink A has $\Omega
= 1.5 \times 10^{-9}$ s$^{-1}$ and sink B has $\Omega = 6 \times
10^{-10}$~s$^{-1}$.  We also have

\begin{equation}
\beta = \frac{\left(1/2\right) I \Omega^2}{q GM^2/R} \mbox{,}
\end{equation}        

\noindent where $I = pMR^2$ is the moment of inertia, $p=2/5$, and
$q=3/5$.  In this case sink A has $\beta = 0.070$ and sink B has
$\beta = 0.044$, very similar to the values derived for the sinks in
\cite{jappsen&klessen2004}.  The requirement for fragmentation ranges
from $\beta > 0.01$ to $\beta > 0.1$, depending on the true density
structure and thermal properties of gas on sub-sink scales, as well as
the effects of magnetic fields (e.g. \citealt{boss&myhill1995,
boss1999}). For the above values of $\beta$, it is thus not entirely
certain whether subfragmentation would occur, and this will need to be
determined with higher resolution studies.

As discussed earlier, however, disk structure is expected even on sub-sink scales, so it would also be appropriate to examine the Toomre criterion for disk fragmentation:

\begin{equation}
Q = \frac{c_{\rmn s} \kappa}{\pi G \Sigma} < 1  \mbox{\ .}
\end{equation}
 
\noindent Here, $\Sigma$ is the disk surface density and $\kappa$ the
epicyclic frequency, which is equal to the angular velocity for a disk
undergoing Keplerian rotation. However, evaluating Q would require
knowledge of disk temperature and surface density on sub-sink scales,
which is not available.  As discussed in \S 3.2, cooling is
likely to occur faster than angular momentum transport, leading to a
sub-sink Keplerian disk.  However, gas within these inner regions is
more susceptible to heating and other protostellar feedback.  This
makes disk fragmentation more likely to occur in the cooler outer
regions of the disk on scales much larger than the sinks, and indeed
such fragmentation is seen in the simulation.  Similar points were
made in studies such as \cite{kratter&matzner2006} and
\cite{krumholzetal2007}.

If a sub-sink binary were to form, however, this could be yet another pathway towards a GRB.  As discussed in studies such as \cite{fryeretal1999}, \cite{bromm&loeb2006}, and \cite{belczynskietal2007}, a binary that is tight enough can allow Roche lobe overflow and a common-envelope phase to occur.  This will remove the hydrogen envelope of the primary, fulfilling one of the requirements for a collapsar GRB.  Even if the stars are rapid rotators, however, \cite{belczynskietal2007} find that Pop III binaries will yield GRBs in only a small fraction, $\la 1 \%$, of cases.  Tidal interactions rarely spin up one of the binary members sufficiently to produce a GRB, and in fact such interactions more often cause the binary members to spin down.     

On the other hand, if a wide 50 AU binary were to form within the sinks, this may still leave enough angular momentum for a GRB to form through the rotational mixing pathway.  The total angular momentum that will go into the binary orbit will be

\begin{equation}
J_{\rmn orb} = M_1 M_2 \frac{\sqrt{aG \left(M_1 + M_2\right) \left(1-e^2\right)}}{M_1 + M_2} \mbox{,} 
\end{equation}              

\noindent where $a$ is the semimajor axis of the orbit, $e$ is the eccentricity, and $M_1$ and $M_2$ are the stellar masses (e.g. \citealt{belczynskietal2007}).  If sink A becomes a circular-orbit binary with $M_1 = M_2 = 17$~M$_{\odot}$, then $J_{\rmn orb} = 5 \times 10^{-2} \, \rmn M_{\odot}$ pc km s$^{-1}$, about $50 \%$ of J$_{\rmn sink}$ for sink A at the end of the simulation.  If the remaining angular momentum went to the spin of each of the binary components, a GRB may still be able to form.

\section{Discussion and Conclusion}

We evolved a three-dimensional SPH cosmological simulation until the
formation of the first minihalo at $z=20$, and then followed the
evolution of the minihalo gas up to maximum density of $10^{12}$
cm$^{-3}$.  After this point we used the sink particle method to
continue the simulation for 5000 years.  A large-scale thick disk of
order 1000 AU that formed around the main sink was resolved and so the
calculation was able to follow angular momentum transport that occurred
within this disk down to resolution length scales of 50 AU.  We find
that there is sufficient angular momentum in Pop~III star-forming
cores, represented by the sink particles, to yield rapidly rotating
Pop III stars.  More specifically, we find that the star-disk systems
are likely to rotate at Keplerian speeds.  This leads to stellar
rotational velocities that can potentially exceed 1000~km s$^{-1}$ for
stars with $M \ga 30 \, {\rmn M}_{\odot}$. This in turn should lead to
chemically homogeneous evolution, yielding hotter and more luminous
stars than without rotation.  The stars should also retain sufficient
spin to power hypernovae as well as collapsar GRBs
(e.g. \citealt{nomotoetal2003, yoon&langer2005, woosley&heger2006}).
Such GRBs may be observed by the {\it Swift} satellite, which has
already detected GRBs at a redshift as high as $z \approx 8.2$
(\citealt{salvaterra2009,tanviretal2009}), and may also be detected by
possible future missions such as JANUS and EXIST.

We emphasize the caveat that we did not fully resolve stellar scales.
We have measured the total angular momentum accreted within $r_{\rmn
acc} \simeq$~50~AU of the star, and we have argued that a Keplerian
disk and perhaps a binary is expected to form on sub-sink scales,
still leaving enough angular momentum for one or two rapidly rotating
stars.  However, there are further processes which can transport
angular momentum away from rotating stars.  For instance, angular
momentum may be lost through stellar winds, but the mass and angular
momentum loss through winds is expected to be much lower for
low-metallicity and Pop III stars than for higher-metallicity stars
(\citealt{nugis&lamers2000, kudritzki2002}).  Other processes include
disk torques induced by gravitational instability as well as viscous
torques, which have a variety of sources including hydromagnetic
instability (see, e.g. \citealt{papa&lin1995} for a review).

In particular, the magneto-hydrodynamic (MHD) aspect of Pop III star
formation is still very uncertain (e.g. \citealt{maki&susa2007}), and
we therefore here neglect any angular momentum loss due to magnetic
torques.  Earlier work, however, gives some hint as to the possible
effect of magnetic fields.  \cite{machidaetal2008} conclude that if a
star-forming primordial cloud has a large enough initial magnetic
field ($B > 10^{-9} \left[n/10^3 \rmn cm^{-3}\right]^{2/3}$ G), a
protostellar jet will be driven provided that the cloud's rotational
energy is less than its magnetic energy.  However, \cite{xuetal2008}
find that the Biermann battery mechanism and flux freezing alone will
not amplify magnetic fields in a collapsing halo quickly enough to
reach this threshold value.  In contrast, small-scale dynamo
amplification as described by \cite{schleicheretal2010} could generate
sufficient magnetic fields for the magneto-rotational instability
(MRI) to operate in primordial protostellar disks
(e.g. \citealt{balbus&hawley1991}).  The resulting turbulent viscosity
would facilitate outward angular momentum transfer in the disk, and it
may further allow generation of sufficient magnetic field strength to
drive collimated protostellar outflows that can also remove angular
momentum (e.g. \citealt{tan&blackman2004, silk&langer2006}).  This may
furthermore facilitate some form of `disk-locking' as described by
various authors such as \cite{koenigl1991}, \cite{shuetal1994}, and
\cite{matt&pudritz2005}, where the stellar rotation will be `locked'
to a rate given by the star's mass, accretion rate, magnetic field
strength, and radius.  Such a model was described by \cite{koenigl1991},
for example, to yield $\Omega_* \sim G M_*^{5/7} \dot{M}^{3/7} B^{-6/7}
R_*^{-18/7} $, where $\Omega_*$ is the star's angular velocity and $B$
the stellar magnetic field strength.
In short, the rate at which these effects will remove angular momentum
from the star is very dependent on the still uncertain magnetic field
strength in Pop III star forming regions, although such effects are a
likely part of the explanation for slowly rotating stars observed in
the Galaxy (see, e.g. \citealt{bodenheimer1995}).  Whether this also
applies in the early Universe will be best determined through future
numerical simulations. A three-dimensional cosmological simulation
that can resolve stellar scales and follow MHD processes for many
dynamical times is highly computationally demanding. For the moment,
our preliminary calculation provides an upper limit for the Pop III
stellar rotation rate.

A comparison with \cite{jappsen&klessen2004} shows interestingly
similar results.  The average specific angular momentum of their
protostellar objects was $8 \times 10^{19}$ cm$^2$ s$^{-1}$, and the
typical mass of each object was $\simeq$ 1 M$_{\odot}$.  They find
that the specific angular momentum increases with mass, with $j
\propto M^{2/3}$ being their preferred fit.  Our sink particles are 9
and 34 times more massive and so should have specific angular momenta
about 4 and 10 times higher, or $3-8 \times 10^{20}$~cm$^2$~s$^{-1}$.
This is indeed the specific angular momentum measured for our sinks.
This also compares well with the range of observed angular momenta of
various structures in the Milky Way.  For instance, the average
specific angular momentum of binaries in the Taurus star-forming
region was found by \cite{simonetal1995} to be $j \simeq$ $2 \times
10^{20}$~cm$^2$~s$^{-1}$, and similar values were found for G-dwarf
stars by \cite{duq&mayor1991}.  For less-evolved structures,
\cite{casellietal2002} found an average of $j = 7 \times
10^{20}$~cm$^2$~s$^{-1}$ for cores of mean mass of 6~M$_{\odot}$.
\cite{goodmanetal1993} observed larger cores of approximately
50~M$_{\odot}$ and obtained an average of $j \simeq$ $2 \times
10^{21}$ cm$^2$ s$^{-1}$.  These cores would be expected to lose
angular momentum as they evolve into protostars, leading to smaller
values similar to those observed in stellar binaries.  Despite the
different initial conditions which give rise to Pop III stars versus
stars in our Galaxy, the overall angular momentum reservoir for both
is very similar.  Thus, just as rapidly rotating massive stars are
observed today (e.g. \citealt{huang&gies2008,wolffetal2008}), rapidly
rotating massive stars seem likely to exist in the early Universe as
well. Unless they are spun down by processes such as magnetic torques
or bipolar outflows, such rapid rotation rates must play an important
role in the evolution and final state of the first stars.

\section*{Acknowledgments}

AS thanks Andreas Pawlik and Milos Milosavljevic for helpful discussions. 
The authors also thank Sylvia Ekstr{\"o}m for valuable feedback.  This work
was supported in part by NSF grants AST-0708795 and AST1009928 and
NASA ATFP grant NNX08AL43G (for VB), and NSF grant AST-0907890 and
NASA grants NNX08AL43G and NNA09DB30A (for AL). The simulations
presented here were carried out at the Texas Advanced Computing Center
(TACC).

\bibliographystyle{mn2e}
\bibliography{sink_am6}{}

\begin{thebibliography}{}

\bibitem[\protect\citeauthoryear{{Abel}, {Bryan} \& {Norman}}{{Abel}
  et~al.}{2002}]{abeletal2002}
{Abel} T.,  {Bryan} G.~L.,    {Norman} M.~L.,  2002, Sci, 295, 93

\bibitem[\protect\citeauthoryear{{Alvarez}, {Bromm} \& {Shapiro}}{{Alvarez}
  et~al.}{2006}]{alvarezetal2006}
{Alvarez} M.~A.,  {Bromm} V.,    {Shapiro} P.~R.,  2006, ApJ, 639, 621

\bibitem[\protect\citeauthoryear{{Balbus} \& {Hawley}}{{Balbus} \&
  {Hawley}}{1991}]{balbus&hawley1991}
{Balbus} S.~A.,  {Hawley} J.~F.,  1991, ApJ, 376, 214

\bibitem[\protect\citeauthoryear{{Barkana} \& {Loeb}}{{Barkana} \&
  {Loeb}}{2001}]{barkana&loeb2001}
{Barkana} R.,  {Loeb} A.,  2001, Phys. Rep., 349, 125

\bibitem[\protect\citeauthoryear{{Bate}, {Bonnell} \& {Price}}{{Bate}
  et~al.}{1995}]{bateetal1995}
{Bate} M.~R.,  {Bonnell} I.~A.,    {Price} N.~M.,  1995, MNRAS, 277, 362

\bibitem[\protect\citeauthoryear{{Bate} \& {Burkert}}{{Bate} \&
  {Burkert}}{1997}]{bate&burkert1997}
{Bate} M.~R.,  {Burkert} A.,  1997, MNRAS, 288, 1060

\bibitem[\protect\citeauthoryear{{Belczynski}, {Bulik}, {Heger} \&
  {Fryer}}{{Belczynski} et~al.}{2007}]{belczynskietal2007}
{Belczynski} K.,  {Bulik} T.,  {Heger} A.,    {Fryer} C.,  2007, ApJ, 664, 986

\bibitem[\protect\citeauthoryear{{Bodenheimer}}{{Bodenheimer}}{1995}]{bodenhei%
mer1995}
{Bodenheimer} P.,  1995, ARA\&A, 33, 199

\bibitem[\protect\citeauthoryear{{Boss}}{{Boss}}{1999}]{boss1999}
{Boss} A.~P.,  1999, ApJ, 520, 744

\bibitem[\protect\citeauthoryear{{Boss} \& {Myhill}}{{Boss} \&
  {Myhill}}{1995}]{boss&myhill1995}
{Boss} A.~P.,  {Myhill} E.~A.,  1995, ApJ, 451, 218

\bibitem[\protect\citeauthoryear{{Bromm}, {Coppi} \& {Larson}}{{Bromm}
  et~al.}{2002}]{brommetal2002}
{Bromm} V.,  {Coppi} P.~S.,    {Larson} R.~B.,  2002, ApJ, 564, 23

\bibitem[\protect\citeauthoryear{{Bromm}, {Ferrara}, {Coppi} \&
  {Larson}}{{Bromm} et~al.}{2001}]{brommetal2001b}
{Bromm} V.,  {Ferrara} A.,  {Coppi} P.~S.,    {Larson} R.~B.,  2001, MNRAS,
  328, 969

\bibitem[\protect\citeauthoryear{{Bromm}, {Kudritzki} \& {Loeb}}{{Bromm}
  et~al.}{2001}]{brommetal2001}
{Bromm} V.,  {Kudritzki} R.~P.,    {Loeb} A.,  2001, ApJ, 552, 464

\bibitem[\protect\citeauthoryear{{Bromm} \& {Larson}}{{Bromm} \&
  {Larson}}{2004}]{bromm&larson2004}
{Bromm} V.,  {Larson} R.~B.,  2004, ARA\&A, 42, 79

\bibitem[\protect\citeauthoryear{{Bromm} \& {Loeb}}{{Bromm} \&
  {Loeb}}{2002}]{bromm&loeb2002}
{Bromm} V.,  {Loeb} A.,  2002, ApJ, 575, 111

\bibitem[\protect\citeauthoryear{{Bromm} \& {Loeb}}{{Bromm} \&
  {Loeb}}{2004}]{bromm&loeb2004}
{Bromm} V.,  {Loeb} A.,  2004, New Astron., 9, 353

\bibitem[\protect\citeauthoryear{{Bromm} \& {Loeb}}{{Bromm} \&
  {Loeb}}{2006}]{bromm&loeb2006}
{Bromm} V.,  {Loeb} A.,  2006, ApJ, 642, 382

\bibitem[\protect\citeauthoryear{{Bromm}, {Yoshida} \& {Hernquist}}{{Bromm}
  et~al.}{2003}]{brommyoshida&hernquist2003}
{Bromm} V.,  {Yoshida} N.,    {Hernquist} L.,  2003, ApJ, 596, L135

\bibitem[\protect\citeauthoryear{{Bromm}, {Yoshida}, {Hernquist} \&
  {McKee}}{{Bromm} et~al.}{2009}]{byhm2009}
{Bromm} V.,  {Yoshida} N.,  {Hernquist} L.,    {McKee} C.~F.,  2009, Nat, 459,
  49

\bibitem[\protect\citeauthoryear{{Caselli}, {Benson}, {Myers} \&
  {Tafalla}}{{Caselli} et~al.}{2002}]{casellietal2002}
{Caselli} P.,  {Benson} P.~J.,  {Myers} P.~C.,    {Tafalla} M.,  2002, ApJ,
  572, 238

\bibitem[\protect\citeauthoryear{{Cesaroni}, {Galli}, {Lodato}, {Walmsley} \&
  {Zhang}}{{Cesaroni} et~al.}{2006}]{cesaronietal2006}
{Cesaroni} R.,  {Galli} D.,  {Lodato} G.,  {Walmsley} M.,    {Zhang} Q.,  2006,
  Nature, 444, 703

\bibitem[\protect\citeauthoryear{{Ciardi} \& {Ferrara}}{{Ciardi} \&
  {Ferrara}}{2005}]{ciardi&ferrara2005}
{Ciardi} B.,  {Ferrara} A.,  2005, SSR, 116, 625

\bibitem[\protect\citeauthoryear{{Clark}, {Glover} \& {Klessen}}{{Clark}
  et~al.}{2008}]{clarketal2008}
{Clark} P.~C.,  {Glover} S.~C.~O.,    {Klessen} R.~S.,  2008, ApJ, 672, 757

\bibitem[\protect\citeauthoryear{{Clark}, {Glover}, {Klessen} \&
  {Bromm}}{{Clark} et~al.}{2010}]{clarketal2010}
{Clark} P.~C.,  {Glover} S.~C.~O.,  {Klessen} R.~S.,    {Bromm} V.,  2010, ApJ,
  submitted (arXiv: 1006.1508)

\bibitem[\protect\citeauthoryear{{Duquennoy} \& {Mayor}}{{Duquennoy} \&
  {Mayor}}{1991}]{duq&mayor1991}
{Duquennoy} A.,  {Mayor} M.,  1991, A\&A, 248, 485

\bibitem[\protect\citeauthoryear{{Ekstr{\"o}m}, {Meynet}, {Chiappini},
  {Hirschi} \& {Maeder}}{{Ekstr{\"o}m} et~al.}{008a}]{ekstrometal2008a}
{Ekstr{\"o}m} S.,  {Meynet} G.,  {Chiappini} C.,  {Hirschi} R.,    {Maeder} A.,
   {2008a}, A\&A, 489, 685

\bibitem[\protect\citeauthoryear{{Ekstr{\"o}m}, {Meynet}, {Maeder} \&
  {Barblan}}{{Ekstr{\"o}m} et~al.}{008b}]{ekstrometal2008b}
{Ekstr{\"o}m} S.,  {Meynet} G.,  {Maeder} A.,    {Barblan} F.,  {2008b}, A\&A,
  478, 467

\bibitem[\protect\citeauthoryear{{Frebel}, {Johnson} \& {Bromm}}{{Frebel}
  et~al.}{2007}]{frebeletal2007}
{Frebel} A.,  {Johnson} J.~L.,    {Bromm} V.,  2007, MNRAS, 380, L40

\bibitem[\protect\citeauthoryear{{Frebel}, {Johnson} \& {Bromm}}{{Frebel}
  et~al.}{2009}]{frebeletal2009}
{Frebel} A.,  {Johnson} J.~L.,    {Bromm} V.,  2009, MNRAS, 392, L50

\bibitem[\protect\citeauthoryear{{Fryer}, {Woosley} \& {Hartmann}}{{Fryer}
  et~al.}{1999}]{fryeretal1999}
{Fryer} C.~L.,  {Woosley} S.~E.,    {Hartmann} D.~H.,  1999, ApJ, 526, 152

\bibitem[\protect\citeauthoryear{{Glover}}{{Glover}}{2005}]{glover2005}
{Glover} S.,  2005, Space Sci. Rev., 117, 445

\bibitem[\protect\citeauthoryear{{Goodman}, {Benson}, {Fuller} \&
  {Myers}}{{Goodman} et~al.}{1993}]{goodmanetal1993}
{Goodman} A.~A.,  {Benson} P.~J.,  {Fuller} G.~A.,    {Myers} P.~C.,  1993,
  ApJ, 406, 528

\bibitem[\protect\citeauthoryear{{Gou}, {M{\'e}sz{\'a}ros}, {Abel} \&
  {Zhang}}{{Gou} et~al.}{2004}]{gouetal2004}
{Gou} L.~J.,  {M{\'e}sz{\'a}ros} P.,  {Abel} T.,    {Zhang} B.,  2004, ApJ,
  604, 508

\bibitem[\protect\citeauthoryear{{Greif}, {Glover}, {Bromm} \&
  {Klessen}}{{Greif} et~al.}{2010}]{greifetal2010}
{Greif} T.~H.,  {Glover} S.~C.~O.,  {Bromm} V.,    {Klessen} R.~S.,  2010, ApJ,
  716, 510

\bibitem[\protect\citeauthoryear{{Greif}, {Johnson}, {Bromm} \&
  {Klessen}}{{Greif} et~al.}{2007}]{greifetal2007}
{Greif} T.~H.,  {Johnson} J.~L.,  {Bromm} V.,    {Klessen} R.~S.,  2007, ApJ,
  670, 1

\bibitem[\protect\citeauthoryear{{Greif}, {Johnson}, {Klessen} \&
  {Bromm}}{{Greif} et~al.}{2009}]{greifetal2009}
{Greif} T.~H.,  {Johnson} J.~L.,  {Klessen} R.~S.,    {Bromm} V.,  2009, MNRAS,
  399, 639

\bibitem[\protect\citeauthoryear{{Haiman}, {Thoul} \& {Loeb}}{{Haiman}
  et~al.}{1996}]{haimanetal1996}
{Haiman} Z.,  {Thoul} A.~A.,    {Loeb} A.,  1996, ApJ, 464, 523

\bibitem[\protect\citeauthoryear{{Heger} \& {Langer}}{{Heger} \&
  {Langer}}{2000}]{heger&langer2000}
{Heger} A.,  {Langer} N.,  2000, ApJ, 544, 1016

\bibitem[\protect\citeauthoryear{{Heger} \& {Woosley}}{{Heger} \&
  {Woosley}}{2002}]{heger&woosley2002}
{Heger} A.,  {Woosley} S.~E.,  2002, ApJ, 567, 532

\bibitem[\protect\citeauthoryear{{Heger}, {Woosley} \& {Spruit}}{{Heger}
  et~al.}{2005}]{hegeretal2005}
{Heger} A.,  {Woosley} S.~E.,    {Spruit} H.~C.,  2005, ApJ, 626, 350

\bibitem[\protect\citeauthoryear{{Huang} \& {Gies}}{{Huang} \&
  {Gies}}{2008}]{huang&gies2008}
{Huang} W.,  {Gies} D.~R.,  2008, ApJ, 683, 1045

\bibitem[\protect\citeauthoryear{{Izzard}, {Ramirez-Ruiz} \& {Tout}}{{Izzard}
  et~al.}{2004}]{izzardetal2004}
{Izzard} R.~G.,  {Ramirez-Ruiz} E.,    {Tout} C.~A.,  2004, MNRaS, 348, 1215

\bibitem[\protect\citeauthoryear{{Jappsen} \& {Klessen}}{{Jappsen} \&
  {Klessen}}{2004}]{jappsen&klessen2004}
{Jappsen} A.,  {Klessen} R.~S.,  2004, A\&A, 423, 1

\bibitem[\protect\citeauthoryear{{Johnson}, {Greif} \& {Bromm}}{{Johnson}
  et~al.}{2007}]{johnsongreif&bromm2007}
{Johnson} J.~L.,  {Greif} T.~H.,    {Bromm} V.,  2007, ApJ, 665, 85

\bibitem[\protect\citeauthoryear{{Kitayama}, {Yoshida}, {Susa} \&
  {Umemura}}{{Kitayama} et~al.}{2004}]{kitayamaetal2004}
{Kitayama} T.,  {Yoshida} N.,  {Susa} H.,    {Umemura} M.,  2004, ApJ, 613, 631

\bibitem[\protect\citeauthoryear{{Koenigl}}{{Koenigl}}{1991}]{koenigl1991}
{Koenigl} A.,  1991, ApJ, 370, L39

\bibitem[\protect\citeauthoryear{{Kratter} \& {Matzner}}{{Kratter} \&
  {Matzner}}{2006}]{kratter&matzner2006}
{Kratter} K.~M.,  {Matzner} C.~D.,  2006, MNRAS, 373, 1563

\bibitem[\protect\citeauthoryear{{Kraus}, {Hofmann}, {Menten}, {Schertl},
  {Weigelt}, {Wyrowski}, {Meilland}, {Perraut}, {Petrov}, {Robbe-Dubois},
  {Schilke} \& {Testi}}{{Kraus} et~al.}{2010}]{krausetal2010}
{Kraus} S.,  {Hofmann} K.,  {Menten} K.~M.,  {Schertl} D.,  {Weigelt} G.,
  {Wyrowski} F.,  {Meilland} A.,  {Perraut} K.,  {Petrov} R.,  {Robbe-Dubois}
  S.,  {Schilke} P.,    {Testi} L.,  2010, Nat, 466, 339

\bibitem[\protect\citeauthoryear{{Krumholz}, {Klein} \& {McKee}}{{Krumholz}
  et~al.}{2007}]{krumholzetal2007}
{Krumholz} M.~R.,  {Klein} R.~I.,    {McKee} C.~F.,  2007, ApJ, 656, 959

\bibitem[\protect\citeauthoryear{{Kudritzki}}{{Kudritzki}}{2002}]{kudritzki200%
2}
{Kudritzki} R.~P.,  2002, ApJ, 577, 389

\bibitem[\protect\citeauthoryear{{Lee}, {Brown} \& {Wijers}}{{Lee}
  et~al.}{2002}]{leeetal2002}
{Lee} C.,  {Brown} G.~E.,    {Wijers} R.~A.~M.~J.,  2002, ApJ, 575, 996

\bibitem[\protect\citeauthoryear{{Lee} \& {Ramirez-Ruiz}}{{Lee} \&
  {Ramirez-Ruiz}}{2006}]{lee&ramirez2006}
{Lee} W.~H.,  {Ramirez-Ruiz} E.,  2006, ApJ, 641, 961

\bibitem[\protect\citeauthoryear{{Loeb}}{{Loeb}}{2010}]{loeb2010}
{Loeb} A.,  2010, {How Did the First Stars and Galaxies Form? Princeton
  University Press, Princeton}

\bibitem[\protect\citeauthoryear{{Machida}, {Matsumoto} \&
  {Inutsuka}}{{Machida} et~al.}{2008}]{machidaetal2008}
{Machida} M.~N.,  {Matsumoto} T.,    {Inutsuka} S.,  2008, ApJ, 685, 690

\bibitem[\protect\citeauthoryear{{Madau}, {Ferrara} \& {Rees}}{{Madau}
  et~al.}{2001}]{madauferrara&rees2001}
{Madau} P.,  {Ferrara} A.,    {Rees} M.~J.,  2001, ApJ, 555, 92

\bibitem[\protect\citeauthoryear{{Maeder}}{{Maeder}}{1987}]{maeder1987}
{Maeder} A.,  1987, A\&A, 178, 159

\bibitem[\protect\citeauthoryear{{Maeder} \& {Meynet}}{{Maeder} \&
  {Meynet}}{2000}]{maeder&meynet2000}
{Maeder} A.,  {Meynet} G.,  2000, ARA\&A, 38, 143

\bibitem[\protect\citeauthoryear{{Maki} \& {Susa}}{{Maki} \&
  {Susa}}{2007}]{maki&susa2007}
{Maki} H.,  {Susa} H.,  2007, PASJ, 59, 787

\bibitem[\protect\citeauthoryear{{Martel}, {Evans} \& {Shapiro}}{{Martel}
  et~al.}{2006}]{marteletal2006}
{Martel} H.,  {Evans} N.~J.,    {Shapiro} P.~R.,  2006, ApJS, 163, 122

\bibitem[\protect\citeauthoryear{{Matt} \& {Pudritz}}{{Matt} \&
  {Pudritz}}{2005}]{matt&pudritz2005}
{Matt} S.,  {Pudritz} R.~E.,  2005, ApJ, 632, L135

\bibitem[\protect\citeauthoryear{{Mori}, {Ferrara} \& {Madau}}{{Mori}
  et~al.}{2002}]{moriferrara&madau2002}
{Mori} M.,  {Ferrara} A.,    {Madau} P.,  2002, ApJ, 571, 40

\bibitem[\protect\citeauthoryear{{Naoz} \& {Bromberg}}{{Naoz} \&
  {Bromberg}}{2007}]{naoz&bromberg2007}
{Naoz} S.,  {Bromberg} O.,  2007, MNRAS, 380, 757

\bibitem[\protect\citeauthoryear{{Narayan} \& {Yi}}{{Narayan} \&
  {Yi}}{1994}]{narayan&yi1994}
{Narayan} R.,  {Yi} I.,  1994, ApJ, 428, L13

\bibitem[\protect\citeauthoryear{{Nomoto}, {Maeda}, {Umeda}, {Ohkubo}, {Deng}
  \& {Mazzali}}{{Nomoto} et~al.}{2003}]{nomotoetal2003}
{Nomoto} K.,  {Maeda} K.,  {Umeda} H.,  {Ohkubo} T.,  {Deng} J.,    {Mazzali}
  P.,  2003, Prog. Theor. Phys. Suppl., 151, 44

\bibitem[\protect\citeauthoryear{{Norman}, {O'Shea} \& {Paschos}}{{Norman}
  et~al.}{2004}]{normanetal2004}
{Norman} M.~L.,  {O'Shea} B.~W.,    {Paschos} P.,  2004, ApJ, 601, L115

\bibitem[\protect\citeauthoryear{{Nugis} \& {Lamers}}{{Nugis} \&
  {Lamers}}{2000}]{nugis&lamers2000}
{Nugis} T.,  {Lamers} H.~J.~G.~L.~M.,  2000, A\&A, 360, 227

\bibitem[\protect\citeauthoryear{{Omukai}}{{Omukai}}{2000}]{omukai2000}
{Omukai} K.,  2000, ApJ, 534, 809

\bibitem[\protect\citeauthoryear{{Omukai} \& {Palla}}{{Omukai} \&
  {Palla}}{2003}]{omukai&palla2003}
{Omukai} K.,  {Palla} F.,  2003, ApJ, 589, 677

\bibitem[\protect\citeauthoryear{{Papaloizou} \& {Lin}}{{Papaloizou} \&
  {Lin}}{1995}]{papa&lin1995}
{Papaloizou} J.~C.~B.,  {Lin} D.~N.~C.,  1995, ARA\&A, 33, 505

\bibitem[\protect\citeauthoryear{{Petrovic}, {Langer}, {Yoon} \&
  {Heger}}{{Petrovic} et~al.}{2005}]{petrovicetal2005}
{Petrovic} J.,  {Langer} N.,  {Yoon} S.,    {Heger} A.,  2005, A\&A, 435, 247

\bibitem[\protect\citeauthoryear{{Salvaterra}, {Della Valle}, {Campana},
  {Chincarini}, {Covino}, {D'Avanzo}, {Fern{\'a}ndez-Soto} \&
  {Guidorzi}}{{Salvaterra} et~al.}{2009}]{salvaterra2009}
{Salvaterra} R.,  {Della Valle} M.,  {Campana} S.,  {Chincarini} G.,  {Covino}
  S.,  {D'Avanzo} P.,  {Fern{\'a}ndez-Soto} A.,    {Guidorzi} C. e.~a.,  2009,
  Nat, 461, 1258

\bibitem[\protect\citeauthoryear{{Schleicher}, {Banerjee}, {Sur}, {Arshakian},
  {Klessen}, {Beck} \& {Spaans}}{{Schleicher}
  et~al.}{2010}]{schleicheretal2010}
{Schleicher} D.~R.~G.,  {Banerjee} R.,  {Sur} S.,  {Arshakian} T.~G.,
  {Klessen} R.~S.,  {Beck} R.,    {Spaans} M.,  2010, A\&A, accepted (arXiv:
  1003.1135)

\bibitem[\protect\citeauthoryear{{Schneider}, {Omukai}, {Inoue} \&
  {Ferrara}}{{Schneider} et~al.}{2006}]{schneideretal2006}
{Schneider} R.,  {Omukai} K.,  {Inoue} A.~K.,    {Ferrara} A.,  2006, MNRAS,
  369, 1437

\bibitem[\protect\citeauthoryear{{Shu}, {Najita}, {Ostriker}, {Wilkin}, {Ruden}
  \& {Lizano}}{{Shu} et~al.}{1994}]{shuetal1994}
{Shu} F.,  {Najita} J.,  {Ostriker} E.,  {Wilkin} F.,  {Ruden} S.,    {Lizano}
  S.,  1994, ApJ, 429, 781

\bibitem[\protect\citeauthoryear{{Silk} \& {Langer}}{{Silk} \&
  {Langer}}{2006}]{silk&langer2006}
{Silk} J.,  {Langer} M.,  2006, MNRAS, 371, 444

\bibitem[\protect\citeauthoryear{{Simon}, {Ghez}, {Leinert}, {Cassar}, {Chen},
  {Howell}, {Jameson}, {Matthews}, {Neugebauer} \& {Richichi}}{{Simon}
  et~al.}{1995}]{simonetal1995}
{Simon} M.,  {Ghez} A.~M.,  {Leinert} C.,  {Cassar} L.,  {Chen} W.~P.,
  {Howell} R.~R.,  {Jameson} R.~F.,  {Matthews} K.,  {Neugebauer} G.,
  {Richichi} A.,  1995, ApJ, 443, 625

\bibitem[\protect\citeauthoryear{{Sokasian}, {Yoshida}, {Abel}, {Hernquist} \&
  {Springel}}{{Sokasian} et~al.}{2004}]{syahs2004}
{Sokasian} A.,  {Yoshida} N.,  {Abel} T.,  {Hernquist} L.,    {Springel} V.,
  2004, MNRAS, 350, 47

\bibitem[\protect\citeauthoryear{{Spitzer}}{{Spitzer}}{1978}]{spitzer1978}
{Spitzer} L.,  1978, {Physical Processes in the Interstellar Medium}.
Wiley, New York

\bibitem[\protect\citeauthoryear{{Springel} \& {Hernquist}}{{Springel} \&
  {Hernquist}}{2002}]{springel&hernquist2002}
{Springel} V.,  {Hernquist} L.,  2002, MNRAS, 333, 649

\bibitem[\protect\citeauthoryear{{Springel}, {Yoshida} \& {White}}{{Springel}
  et~al.}{2001}]{springeletal2001}
{Springel} V.,  {Yoshida} N.,    {White} S.~D.~M.,  2001, New Astron., 6, 79

\bibitem[\protect\citeauthoryear{{Spruit}}{{Spruit}}{2002}]{spruit2002}
{Spruit} H.~C.,  2002, A\&A, 381, 923

\bibitem[\protect\citeauthoryear{{Stacy}, {Greif} \& {Bromm}}{{Stacy}
  et~al.}{2010}]{stacyetal2010}
{Stacy} A.,  {Greif} T.~H.,    {Bromm} V.,  2010, MNRAS, 403, 45

\bibitem[\protect\citeauthoryear{{Stahler}, {Palla} \& {Salpeter}}{{Stahler}
  et~al.}{1986}]{stahler&palla1986}
{Stahler} S.~W.,  {Palla} F.,    {Salpeter} E.~E.,  1986, ApJ, 302, 590

\bibitem[\protect\citeauthoryear{{Suwa} \& {Ioka}}{{Suwa} \&
  {Ioka}}{2010}]{suwa&ioka2010}
{Suwa} Y.,  {Ioka} K.,  2010, arXiv:1009.6001

\bibitem[\protect\citeauthoryear{{Tan} \& {Blackman}}{{Tan} \&
  {Blackman}}{2004}]{tan&blackman2004}
{Tan} J.~C.,  {Blackman} E.~G.,  2004, ApJ, 603, 401

\bibitem[\protect\citeauthoryear{{Tan} \& {McKee}}{{Tan} \&
  {McKee}}{2004}]{tan&mckee2004}
{Tan} J.~C.,  {McKee} C.~F.,  2004, ApJ, 603, 383

\bibitem[\protect\citeauthoryear{{Tanvir}, {Fox}, {Levan}, {Berger},
  {Wiersema}, {Fynbo}, {Cucchiara}, {Kr{\"u}hler}, {Gehrels}, {Bloom} \&
  {Greiner}}{{Tanvir} et~al.}{2009}]{tanviretal2009}
{Tanvir} N.~R.,  {Fox} D.~B.,  {Levan} A.~J.,  {Berger} E.,  {Wiersema} K.,
  {Fynbo} J.~P.~U.,  {Cucchiara} A.,  {Kr{\"u}hler} T.,  {Gehrels} N.,  {Bloom}
  J.~S.,    {Greiner} J. e.~a.,  2009, Nat, 461, 1254

\bibitem[\protect\citeauthoryear{{Tegmark}, {Silk}, {Rees}, {Blanchard}, {Abel}
  \& {Palla}}{{Tegmark} et~al.}{1997}]{tegmarketal1997}
{Tegmark} M.,  {Silk} J.,  {Rees} M.~J.,  {Blanchard} A.,  {Abel} T.,
  {Palla} F.,  1997, ApJ, 474, 1

\bibitem[\protect\citeauthoryear{{Tornatore}, {Ferrara} \&
  {Schneider}}{{Tornatore} et~al.}{2007}]{tfs2007}
{Tornatore} L.,  {Ferrara} A.,    {Schneider} R.,  2007, MNRAS, 382, 945

\bibitem[\protect\citeauthoryear{{Turk}, {Abel} \& {O'Shea}}{{Turk}
  et~al.}{2009}]{turketal2009}
{Turk} M.~J.,  {Abel} T.,    {O'Shea} B.,  2009, Sci, 325, 601

\bibitem[\protect\citeauthoryear{{Ulrich}}{{Ulrich}}{1976}]{ulrich1976}
{Ulrich} R.~K.,  1976, ApJ, 210, 377

\bibitem[\protect\citeauthoryear{{Wada} \& {Venkatesan}}{{Wada} \&
  {Venkatesan}}{2003}]{wada&venkatesan2003}
{Wada} K.,  {Venkatesan} A.,  2003, ApJ, 591, 38

\bibitem[\protect\citeauthoryear{{Whalen}, {Abel} \& {Norman}}{{Whalen}
  et~al.}{2004}]{whalenetal2004}
{Whalen} D.,  {Abel} T.,    {Norman} M.~L.,  2004, ApJ, 610, 14

\bibitem[\protect\citeauthoryear{{Wise} \& {Abel}}{{Wise} \&
  {Abel}}{2008}]{wise&abel2008}
{Wise} J.~H.,  {Abel} T.,  2008, ApJ, 685, 40

\bibitem[\protect\citeauthoryear{{Wolff}, {Strom}, {Cunha}, {Daflon}, {Olsen}
  \& {Dror}}{{Wolff} et~al.}{2008}]{wolffetal2008}
{Wolff} S.~C.,  {Strom} S.~E.,  {Cunha} K.,  {Daflon} S.,  {Olsen} K.,
  {Dror} D.,  2008, AJ, 136, 1049

\bibitem[\protect\citeauthoryear{{Woosley}}{{Woosley}}{1993}]{woosley1993}
{Woosley} S.~E.,  1993, ApJ, 405, 273

\bibitem[\protect\citeauthoryear{{Woosley} \& {Bloom}}{{Woosley} \&
  {Bloom}}{2006}]{woosley&bloom2006}
{Woosley} S.~E.,  {Bloom} J.~S.,  2006, ARA\&A, 44, 507

\bibitem[\protect\citeauthoryear{{Woosley} \& {Heger}}{{Woosley} \&
  {Heger}}{2006}]{woosley&heger2006}
{Woosley} S.~E.,  {Heger} A.,  2006, ApJ, 637, 914

\bibitem[\protect\citeauthoryear{{Xu}, {O'Shea}, {Collins}, {Norman}, {Li} \&
  {Li}}{{Xu} et~al.}{2008}]{xuetal2008}
{Xu} H.,  {O'Shea} B.~W.,  {Collins} D.~C.,  {Norman} M.~L.,  {Li} H.,    {Li}
  S.,  2008, ApJ, 688, L57

\bibitem[\protect\citeauthoryear{{Yoon} \& {Langer}}{{Yoon} \&
  {Langer}}{2005}]{yoon&langer2005}
{Yoon} S.,  {Langer} N.,  2005, A\&A, 443, 643

\bibitem[\protect\citeauthoryear{{Yoshida}, {Abel}, {Hernquist} \&
  {Sugiyama}}{{Yoshida} et~al.}{2003}]{yahs2003}
{Yoshida} N.,  {Abel} T.,  {Hernquist} L.,    {Sugiyama} N.,  2003, ApJ, 592,
  645

\bibitem[\protect\citeauthoryear{{Yoshida}, {Omukai}, {Hernquist} \&
  {Abel}}{{Yoshida} et~al.}{2006}]{yoshidaetal2006}
{Yoshida} N.,  {Omukai} K.,  {Hernquist} L.,    {Abel} T.,  2006, ApJ, 652, 6

\bibitem[\protect\citeauthoryear{{Zhang}, {Woosley} \& {Heger}}{{Zhang}
  et~al.}{2004}]{zhangetal2004}
{Zhang} W.,  {Woosley} S.~E.,    {Heger} A.,  2004, ApJ, 608, 365

\end{thebibliography}

\label{lastpage}

\end{document}